\documentclass[sort&compress,final,3p,times,twocolumn,fixfloat]{elsarticle}
\usepackage{epsfig,slashed,nicefrac}
\usepackage{graphicx}
\usepackage[utf8]{inputenc}
\usepackage{color}
\usepackage{multirow}
\usepackage{float} 
\usepackage{subfig} 
\usepackage{eqnarray,amsmath}
\usepackage{paralist}
\usepackage{times}
\usepackage{txfonts}
\usepackage{comment}
\usepackage[normalem]{ulem}

\journal{Physics Letters B}

\newcommand{\eg}{{\it e.g.}}
\newcommand{\ie}{{\it i.e.}}

\newcommand{\ho}{{\sc HELAC-Onia}}

\newcommand{\gammaupc}{{\sc gamma-UPC}}
\newcommand{\superchic}{{\sc SuperChic}}
\newcommand{\mgamc}{{\sc MadGraph5\_aMC@NLO}}
\newcommand{\qgraf}{{\sc Qgraf}}
\newcommand{\feynarts}{{\sc FeynArts}}
\newcommand{\finiteflow}{{\sc FiniteFlow}}
\newcommand{\kira}{{\sc Kira}}
\newcommand{\multivariateapart}{{\sc MultivariateApart}}

\newcommand{\lambdavec}{\vec{\lambda}}

\def\be{\begin{equation*}}
\def\ee{\end{equation*}}
\def\bsp#1\esp{\begin{split}#1\end{split}} 
\def\bpm{\begin{pmatrix}}
\def\epm{\end{pmatrix}}

 \usepackage{ifpdf}
 \ifpdf
 \usepackage[pdftex]{hyperref}
 \else
 \usepackage[hypertex]{hyperref}
 \fi

 \hypersetup{
   pdftitle={},%
   pdfauthor={},%
   pdfsubject={},%
   pdfkeywords={},%
   pdfstartview={},%
   bookmarksopen=true, breaklinks=true, debug=true, %
   colorlinks=true, linkcolor=red, citecolor=blue, urlcolor=blue
 }

\newcount\savefnused
\newcount\savefndone

\newcommand{\savefootnote}[2][\empty]%
{\ifx\empty#1\footnotemark\else\footnotemark[#1]\fi
 \global\advance\savefnused by 1
 \expandafter\xdef\csname savefnmark\the\savefnused\endcsname{\thefootnote}%
 \expandafter\xdef\csname savefntext\the\savefnused\endcsname{#2}%
}
\newcommand{\flushfootnote}{\loop\ifnum\savefndone<\savefnused
  \global\advance\savefndone by 1
  \footnotetext[\csname savefnmark\the\savefndone\endcsname]%
    {\csname savefntext\the\savefndone\endcsname}%
  \global\expandafter\let\csname savefnmark\the\savefndone\endcsname\relax
  \global\expandafter\let\csname savefntext\the\savefndone\endcsname\relax
\repeat}

\begin{document}

\title{Light-by-Light Scattering at Next-to-Leading Order in QCD and QED}

\address[LPTHE]{Laboratoire de Physique Th\'eorique et Hautes Energies (LPTHE), UMR 7589, Sorbonne Universit\'e et CNRS, 4 place Jussieu, 75252 Paris Cedex 05, France}
\address[Bonn]{Bethe Center for Theoretical Physics, Universit\"at Bonn, D-53115, Germany}
\address[Bern]{Institute for Theoretical Physics, University of Bern, Sidlerstrasse 5, 3012 Bern, Switzerland}

\author[LPTHE]{Ajjath~A.~H.~}
\ead{aabdulhameed@lpthe.jussieu.fr}

\author[Bonn]{Ekta Chaubey}
\ead{eekta@uni-bonn.de}

\author[Bern]{Mathijs Fraaije}
\ead{mathijsfraaije@gmail.com}

\author[Bern]{Valentin Hirschi}
\ead{valentin.hirschi@gmail.com}

\author[LPTHE]{Hua-Sheng Shao}
\ead{huasheng.shao@lpthe.jussieu.fr}

\date{\today}

\begin{abstract}
The recent experimental observation of Light-by-Light (LbL) scattering at the Large Hadron Collider has revived interest in this fundamental process, and especially of the accurate prediction of its cross-section, which we present here for the first time at Next-to-Leading Order (NLO) in both QCD and QED. We compare two radically different computational approaches, both exact in the fermion mass dependence, thus offering a strong cross-check of our results. The first approach is a fully analytic method to calculate compact and well-organized two-loop helicity amplitudes. The second one is entirely numerical and leverages the Local Unitarity construction. Our two calculations agree with each other and conclude that including the exact fermion mass contribution typically increases the size of the NLO corrections. Moreover, we find that the exact result converges slowly to the massless limit of the high-energy regime, thus emphasizing the importance of including the full mass dependence at NLO. We also compare our results with the ATLAS measurement of LbL in ultra-peripheral lead-lead collisions, and find that the inclusion of exact NLO corrections reduces, but does not eliminate, the existing tension with theoretical predictions.
\end{abstract}

\maketitle

\section{\label{sec:intro}Introduction}

Light, a form of ElectroMagnetic (EM) radiation, has revolutionized our understanding of Nature, \eg, from Maxwell's equations to the establishments of the two pillars of modern physics -- Einstein's theory of relativity and quantum mechanics. A natural question about light that needs answering is how it interacts with itself. The abelian nature of Quantum ElectroDynamics (QED) prevents any direct interaction, and due to Furry's theorem~\cite{Furry:1937zz}, the three-photon interaction vanishes in QED and is strongly suppressed in ElectroWeak (EW) theory~\cite{Delbourgo:1976gt,Basham:1977rj}. Therefore, the four-photon vertex is the most viable light self-interaction, and it can be directly probed by studying the Light-by-Light (LbL) scattering process $\gamma\gamma\to\gamma\gamma$. LbL scattering predictions date back to the  1930s~\cite{Heisenberg:1934pza,Euler:1935zz,Euler:1935qgl,Heisenberg:1936nmg}. However, its experimental evidence remained elusive and only recently emerged at the Large Hadron Collider (LHC) through Ultra-Peripheral heavy-ion Collisions (UPCs)~\cite{ATLAS:2017fur,CMS:2018erd,ATLAS:2019azn,ATLAS:2020hii}, where the impact parameter is larger than the sum of the two ion radii and the ions scatter quasi-elastically. The experimental feasibility of observing LbL has revived broad theoretical interest in studying the process in the context of both the Standard Model (SM) and beyond the SM (BSM) of particle physics. In the literature, LbL has been suggested to be an ideal probe of the bound states of leptons~\cite{dEnterria:2022ysg,dEnterria:2023yao} and gluons~\cite{Greiner:1992fz}, the quartic anomalous gauge couplings~\cite{dEnterria:2013zqi}, the axion-like particles~\cite{Knapen:2016moh}, the graviton-like particles~\cite{dEnterria:2023npy,Atag:2010bh}, the nonlinear Born-Infeld extensions of QED and SM~\cite{Ellis:2017edi}, the photon self-interaction from the noncommutative QED~\cite{Horvat:2020ycy}, large extra dimensions~\cite{Cheung:1999ja,Davoudiasl:1999di}, and supersymmetry~\cite{Greiner:1992fz}.

The \textit{direct} experimental observation of $\gamma\gamma\to\gamma\gamma$ was only achieved recently in heavy-ion UPCs at the LHC~\cite{ATLAS:2017fur,CMS:2018erd,ATLAS:2019azn,ATLAS:2020hii}, thanks to the large coherent photon flux carried by the ultra-relativistic
nucleus with a large charge number $Z$ (\eg, the lead Pb ion has $Z=82$). The cross-section in lead-lead (Pb-Pb) collisions benefits from a huge enhancement factor of $Z^4\approx 4.5\cdot 10^7$ compared to analogous proton-proton and electron-positron collisions. The LbL measurement is carried out by exclusively reconstructing all final particles, while the initial ions remain intact~\footnote{Strictly speaking, the ions do not need to be intact since they can be EM excited and emit neutrons. These forward neutrons can be tagged. However, in this paper, we only consider the inclusive case with respect to the forward neutrons. Therefore, the word ``intact" here should be understood in a loose sense.}. In contrast to more inclusive analyses, the exclusive nature of the final state definition allows one to efficiently reject background contributions, loosen experimental triggers, and access new kinematic regimes at the LHC. Such exclusive events can reliably be identified with detectors at a very forward angle, such as Zero Degree Calorimeters in nucleus-nucleus collisions.

From the theoretical point of view, LbL scattering is a loop-induced process that is mediated at the lowest order by one-loop charged fermions and $W^\pm$ boson Feynman diagrams in the SM, with its Leading-Order (LO) being $\mathcal{O}(\alpha^4)$, where $\alpha$ is the EM fine-structure constant. In the Low-Energy (LE) limit, the process can be effectively described by the Euler-Heisenberg Lagrangian~\cite{Heisenberg:1936nmg}, which has been computed up to next-to-leading order (NLO)~\cite{Martin:2003gb}. On the other hand, the massless limit of the fermion loops has been known at NLO in both Quantum ChromoDynamics (QCD) and QED for two decades~\cite{Bern:2001dg,Binoth:2002xg}. The existing calculations of the cross-section by two Monte Carlo (MC) event generators, \superchic~\cite{Harland-Lang:2020veo} and \gammaupc+\mgamc~\cite{Shao:2022cly,Alwall:2014hca}, based on the one-loop matrix elements, reveal that the ATLAS measurement~\cite{ATLAS:2020hii} in Pb-Pb UPCs is about $2\sigmaup$ above the theory predictions for the fiducial cross-section. To better understand this discrepancy, it is important to study NLO quantum corrections. Since the energy scale of the ATLAS measurement ranges from a few GeV to tens of GeV, the massless limit does not apply to the tau lepton and heavy-flavor quark loops. It is therefore necessary to carry out two-loop computations retaining the exact dependence in the fermion masses to reliably assess the importance of NLO corrections, which is the main motivation of this letter.

In this letter, we present for the first time the fully differential cross-section of the LbL process $\gamma\gamma\to\gamma\gamma$ at NLO accuracy in QCD and QED, with exact dependence in the fermion masses. The computation is carried out twice using two radically different and independent approaches. The first one is fully analytic, based on canonical differential equation systems for solving two-loop master integrals and integration-by-parts reduction for obtaining compact two-loop helicity amplitudes. The second approach considers the Local Unitarity (LU) construction~\cite{Capatti:2020xjc,Capatti:2022tit} to directly compute the fully differential cross-section entirely numerically. We have cross-checked the results from these two methods against each other and found perfect agreement. We stress that this is the first time that LU, or any fully numerical method in momentum space, yields results for a yet unknown cross-section. Our results bring the accuracy of the theoretical prediction for the LbL cross-section down to the percent-level in the regime where the di-photon invariant mass ranges from a few GeV up to the $W^\pm$ mass threshold $2m_W$.

\section{Methodology}

As already mentioned, our first approach is based on the derivation of the analytic one- and two-loop helicity amplitudes for the process $\gamma(p_1,\lambda_1)+\gamma(p_2,\lambda_2)+\gamma(p_3,\lambda_3)+\gamma(p_4,\lambda_4)\to 0$, where all external four momenta $p_i$ are incoming and $\lambda_i$'s are the helicities of the photons. Up to any loop order in QCD and QED, we can decompose the helicity amplitude onto the following tensor basis:
\begin{equation}
\begin{aligned}
\mathcal{M}_{\lambdavec}&=\left(\prod_{i=1}^{4}{\varepsilon_{\lambda_i,\mu_i}(p_i)} \right)\mathcal{M}^{\mu_1 \mu_2 \mu_3 \mu_4}(p_1, p_2, p_3, p_4),\\
\mathcal{M}^{\mu_1 \mu_2 \mu_3 \mu_4}&=A_1 g^{\mu_1 \mu_2} g^{\mu_3 \mu_4} +A_2 g^{\mu_1 \mu_3} g^{\mu_2 \mu_4} + A_3 g^{\mu_1 \mu_4} g^{\mu_2 \mu_3} \\
&\quad+ \sum_{j_1, j_2=1}^3 \Big(B_{j_1 j_2}^1 g^{\mu_1 \mu_2} p_{j_1}^{\mu_3} p_{j_2}^{\mu_4} + B^2_{j_1j_2} g^{\mu_1 \mu_3} p_{j_1}^{\mu_2} p_{j_2}^{\mu_4} \\
&\quad+B^3_{j_1j_2} g^{\mu_1 \mu_4} p_{j_1}^{\mu_2} p_{j_2}^{\mu_3} +B^4_{j_1j_2} g^{\mu_2 \mu_3} p_{j_1}^{\mu_1} p_{j_2}^{\mu_4} \\
&\quad+B^5_{j_1j_2} g^{\mu_2 \mu_4} p_{j_1}^{\mu_1} p_{j_2}^{\mu_3}+B^6_{j_1j_2} g^{\mu_3 \mu_4} p_{j_1}^{\mu_1} p_{j_2}^{\mu_2}\Big) \\
&\quad+\sum_{j_1, j_2, j_3, j_4=1}^3 C_{j_1 j_2 j_3 j_4} p_{j_1}^{\mu_1} p_{j_2}^{\mu_2} p_{j_3}^{\mu_3} p_{j_4}^{\mu_4},\label{eq:helamp}
\end{aligned}
\end{equation}
where $\lambdavec=(\lambda_1,\lambda_2,\lambda_3,\lambda_4)$, $\varepsilon_{\lambda_i,\mu_i}$ are the photon polarization vectors, and the coefficients $A_i$, $B_{jk}^i$ and $C_{ijkl}$ are functions of the Mandelstam variables $s=(p_1+p_2)^2,\; t=(p_2+p_3)^2$, and $u=(p_1+p_3)^2$, as well as the masses of the internal fermions. Taking advantage of the transversality of external photons, Bose symmetry and gauge invariance makes it possible to rewrite the original 138 form factors into only $7$ independent ones $A_1(s,t,u),A_1(t,s,u),A_1(u,s,t),\Delta B_{11}^{1}(s,t,u),\Delta B_{11}^{1}(t,s,u)$,
$\Delta B_{11}^{1}(u,s,t)$, and $\Delta C_{2111}(s,t,u)$~\cite{H:2023wfg}, where we have defined the form factor differences $\Delta B^{1}_{11}(s,t,u)=B^1_{11}(s,t,u)-B^1_{12}(s,t,u)$, and $\Delta C_{2111}(s,t,u)=C_{2111}(s,t,u)-C_{2112}(s,t,u)$. The original form factors in eq.(\ref{eq:helamp}), all with implicit arguments $(s,t,u)$, have also been rewritten into form factors with permutations of the arguments. The $5$ independent helicity amplitudes only depend on the following $5$ form factors $A_S(s,t,u)$, $\Delta \hat{B}_{11}^{1}(s,t,u),\;\Delta \hat{B}_{11}^{1}(t,s,u),\;\Delta \hat{B}_{11}^{1}(u,s,t)$, and $\Delta \hat{C}_{2111}(s,t,u)$ as follows:
\begin{equation}
\begin{aligned}
{\footnotesize
\left(\begin{array}{c}
\mathcal{M}_{++++}\\
\mathcal{M}_{-+++}\\
\mathcal{M}_{--++}\\
\mathcal{M}_{+-+-}\\
\mathcal{M}_{+--+}
\end{array}\right)=\frac{1}{4}\left(\begin{array}{ccccc}
1 & 2 & 2 & 2 & -1 \\
1 & 0 & 0 & 0 & 1 \\
1 & 2 & -2 & -2 & -1 \\
1 & -2 & -2 & 2 & -1 \\
1 & -2 & 2 & -2 & -1 \\
\end{array}\right)\left(\begin{array}{c}
A_S(s,t,u)\\
u\Delta \hat{B}^1_{11}(s,t,u)\\
s\Delta \hat{B}^1_{11}(t,u,s)\\
t\Delta \hat{B}^1_{11}(u,s,t)\\
su\Delta \hat{C}_{2111}(s,t,u)
\end{array}\right),
}
\end{aligned}
\end{equation}
where $A_S(s,t,u)=A_1(s,t,u)+A_1(t,s,u)+A_1(u,s,t)$, $\Delta \hat{B}^{1}_{11}(s,t,u)=\Delta B^{1}_{11}(s,t,u)+A_1(s,t,u)/u$, and $\Delta \hat{C}_{2111}(s,t,u)=\Delta C_{2111}(s,t,u)-A_S(s,t,u)/(su)$. The form factors $A_S(s,t,u)$, $su\Delta \hat{C}_{2111}(s,t,u)$, and the all-plus and one-minus amplitudes $\mathcal{M}_{++++}$ and $\mathcal{M}_{-+++}$ are fully symmetric in $s,t,u$. The last three amplitudes can be related to each other using crossing symmetry: $\mathcal{M}_{+-+-}=\left.\mathcal{M}_{--++}\right|_{s\leftrightarrow u}=\left.\mathcal{M}_{+--+}\right|_{t\leftrightarrow u}$.
All other helicity amplitudes can be obtained from these $5$ independent ones using charge conjugation $C$, parity $P$, and time reversal $T$ symmetries.

Our work only considers the two-loop QCD and QED amplitudes mediated through a fermion loop. The two-loop contribution mediated by the $W^\pm$ boson would require the inclusion of the complete set of EW corrections, which is beyond the scope of this letter.

\subsection{Analytical approach}

For a given fermion species, the two-loop amplitudes, generated with \qgraf~\cite{Nogueira:1991ex} and \feynarts~\cite{Hahn:2000kx}, are reduced to a linear combination of master integrals using integration-by-parts identities as implemented in \finiteflow~\cite{Peraro:2019svx} and \kira~\cite{Maierhofer:2017gsa,Klappert:2020nbg}, in $d=4-2\epsilon$ dimensions. In order to solve the master integrals analytically, we transform them into a canonical basis~\cite{Caron-Huot:2014lda} that satisfies the $\epsilon$-form of differential equations~\cite{Henn:2013pwa} with uniform transcendental weight. The solution can then be generally expressed through iterated integrals, that can be expressed either as multiple polylogarithms or retained as iterated integrals with dlog one-forms. Both of these representations allow fast numerical evaluations. Some of these integrals have also been studied in the context of other processes~\cite{Caron-Huot:2014lda,Xu:2018eos,Mandal:2018cdj,Maltoni:2018zvp,Wang:2020nnr}. In addition, to simplify the final expression of the helicity amplitudes, we use the shuffle algebra of iterated integrals and the symbol techniques~\cite{Duhr:2011zq,Duhr:2019tlz} to find relations among master integrals at a given transcendental weight. The rational coefficients are further simplified using \finiteflow\ and \multivariateapart~\cite{Heller:2021qkz} as well as taking advantage of the symmetry properties of the amplitudes. The final two-loop amplitudes renormalized in the On-Shell (OS) scheme are short enough to be written in a few pages~\cite{H:2023wfg}, which allows us to analytically verify pole cancellations, and also compute interesting kinematic limits.

\subsection{Numerical approach}

\begin{figure}[hbt!]
\includegraphics[width=0.75\columnwidth,draft=false]{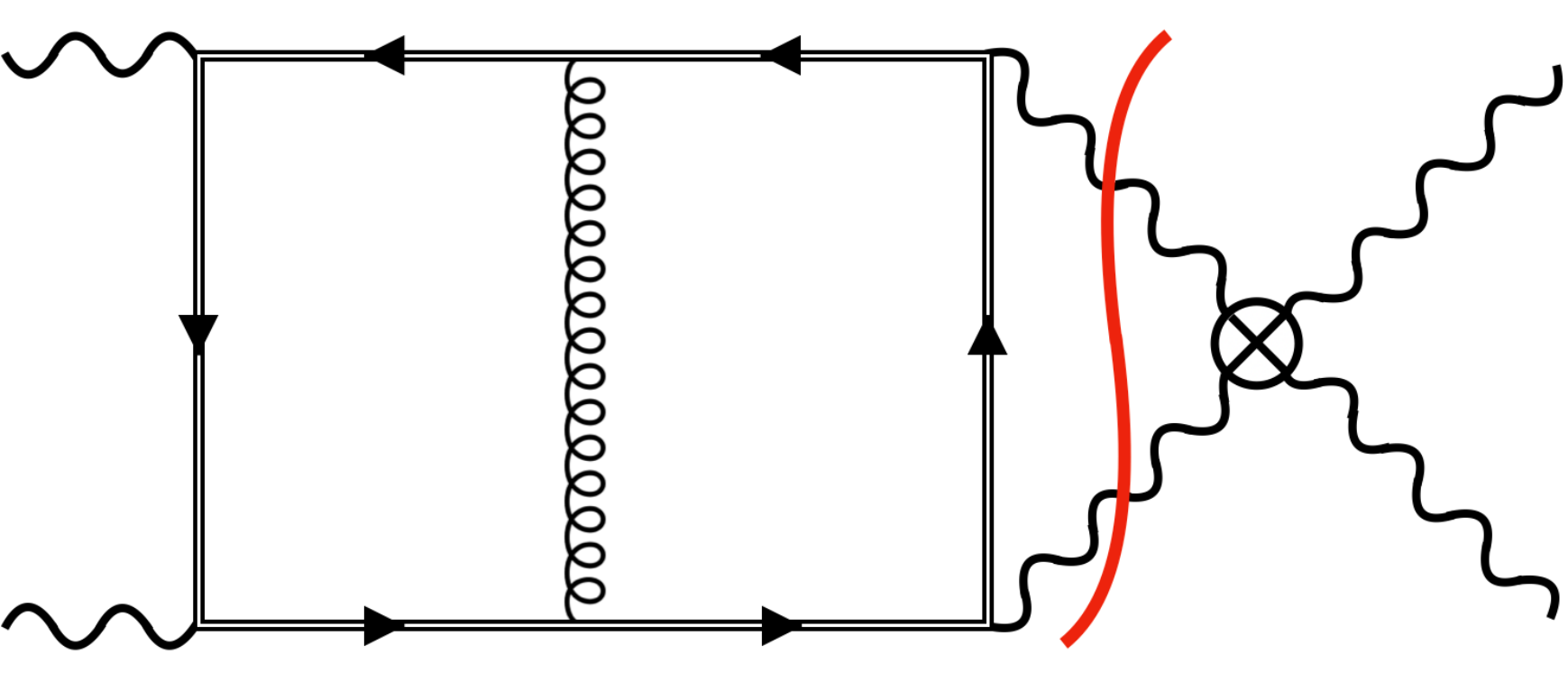}
\caption{Example of one of the 16 distinct 3-loop FSG contributing to the NLO correction of the LbL cross-section. The single Cutkosky cut contributing is shown in red. The effective four-photon vertex is denoted with a cross and is implemented with the exact 1-loop amplitude. The double line corresponds to a massive fermion.}
\label{figLU} \vspace*{0.0cm}
\end{figure}

Our second computational approach relies on direct MC integration in momentum space using the LU construction. The starting point of this method is to consider all Forward-Scattering Graphs (FSG) relevant to a particular cross-section, and for each of them collect all Cutkosky cuts contributing to the scattering process definition of interest.
For the LbL process, each FSG contains a single Cutkosky cut traversing two photons and separating the FSG into a two-loop (at NLO) diagram on the left of the cut, and a one-loop diagram on the right of the cut, thus forming one particular interference term contributing to $2\Re{\left\{\mathcal{M}^{\text{2-loop}}\mathcal{M}^{\text{1-loop}\;\star}\right\}}$. 
To simplify our computation, we chose to write the 1-loop amplitude as an effective four-photon vertex with the exact 1-loop amplitude $\mathcal{M}^{\mu_1 \mu_2 \mu_3 \mu_4}$ as its Feynman rule. At LO (NLO), this leads to 2 (16) distinct non-isomorphic two-(three-)loop FSGs, see fig.~\ref{figLU} for an example FSG contributing at NLO.
\vspace{-0.0cm}

Loop integrals are computed numerically, together with the phase-space integral stemming from the integration of the momenta entering the Cutkosky cut and accompanied by an observable function. LU achieves this by projecting momenta onto the Cutkosky cut using the causal flow~\cite{Capatti:2020xjc} and by writing loop integrals using the Loop-Tree Duality (LTD)~\cite{Catani:2008xa,Bierenbaum:2010cy,Runkel:2019yrs,Runkel:2019zbm,Capatti:2019ypt,Tomboulis:2017rvd,Capatti:2022mly} expression, obtained by analytic integration over the energy components of the loop momenta. In this work, the manifestly causal LTD expression~\cite{Capatti:2020ytd} was used. Ultra-Violet (UV) singularities are locally subtracted in a way that immediately yields results renormalized in the OS scheme~\cite{Capatti:2022tit}. Threshold singularities appear when $\sqrt{s} > 2 m_f$ and they are regularized using a contour deformation of the spatial loop momenta~\cite{Capatti:2019edf}. To mitigate large gauge cancellations between FSGs, we wrote the photon spin-sums in the temporal axial gauge with $n = (1, 0, 0, 0)$.
The resulting 9-dimensional momentum-space integral is then evaluated using MC methods with importance sampling.

\section{Results}

We first set the notation used in the presentation of our results. For a given fermion $f$ with mass $m_f$ we denote the one- and two-loop QCD and QED amplitudes as $\mathcal{M}_{\lambdavec}^{(0,0,f)}$, $\mathcal{M}_{\lambdavec}^{(1,0,f)}$ and $\mathcal{M}_{\lambdavec}^{(0,1,f)}$ respectively. For the $W^\pm$ boson, since we only consider the one-loop result, we denote its amplitude as $\mathcal{M}^{(0,0,W)}_{\lambdavec}$. The amplitude combinations that will be used in the cross-sections are then $\mathcal{M}^{(0,0)}_{\lambdavec}=\sum_{l=f,W}{\mathcal{M}^{(0,0,l)}_{\lambdavec}}$, $\mathcal{M}^{(i,j)}_{\lambdavec}=\sum_{f}{\mathcal{M}^{(i,j,f)}_{\lambdavec}}$ for $(i,j)=(1,0),(0,1)$, $\mathcal{M}^{(1,1,f)}_{\lambdavec}=\mathcal{M}^{(1,0,f)}_{\lambdavec}+\mathcal{M}^{(0,1,f)}_{\lambdavec}$, and
$\mathcal{M}^{(1,1)}_{\lambdavec}=\sum_{f}{\mathcal{M}^{(1,1,f)}_{\lambdavec}}$.
Then, the LO partonic cross-sections can be computed with
\begin{equation}
\begin{aligned}
\hat{\sigma}^{(0,0,l)}&=\frac{1}{2s}\int{{\rm d}\Phi_2\overline{\sum}_{\rm helicity}{\left|\mathcal{M}_{\lambdavec}^{(0,0,l)}\right|^2}},\quad l=f,W,\\
\hat{\sigma}^{(0,0)}&=\frac{1}{2s}\int{{\rm d}\Phi_2\overline{\sum}_{\rm helicity}{\left|\mathcal{M}_{\lambdavec}^{(0,0)}\right|^2}},\label{eq:xsdef0}
\end{aligned}
\end{equation}
where ${\rm d}\Phi_2$ is the 2-body phase-space measure, and the overlined sum denotes the sum over the helicity configurations of the external photons and the average over the initial photon polarizations. We can also define the NLO QCD and/or QED corrections to the partonic cross-sections as follows
\begin{equation}
\begin{aligned}
\hat{\sigma}^{(i,j,f)}&=\frac{1}{2s}\int{{\rm d}\Phi_2\overline{\sum}_{\rm helicity}{2\Re{\left\{\mathcal{M}_{\lambdavec}^{(0,0,f)\;\star} \mathcal{M}_{\lambdavec}^{(i,j,f)}\right\}}}},\\
\hat{\sigma}^{(i,j)}&=\frac{1}{2s}\int{{\rm d}\Phi_2\overline{\sum}_{\rm helicity}{2\Re{\left\{\mathcal{M}_{\lambdavec}^{(0,0)\;\star} \mathcal{M}_{\lambdavec}^{(i,j)}\right\}}}}\label{eq:xsdef1}
\end{aligned}
\end{equation}
with $(i,j)\in \left\{(1,0),(0,1),(1,1)\right\}$. Thus, the NLO QCD, NLO QED, and NLO QCD+QED cross-sections are $\hat{\sigma}^{{\rm NLO}_{\rm QCD}}=\hat{\sigma}^{(0,0)}+\hat{\sigma}^{(1,0)}$, $\hat{\sigma}^{{\rm NLO}_{\rm QED}}=\hat{\sigma}^{(0,0)}+\hat{\sigma}^{(0,1)}$, and $\hat{\sigma}^{{\rm NLO}_{\rm QCD+QED}}=\hat{\sigma}^{(0,0)}+\hat{\sigma}^{(1,1)}$ respectively. We have analogous definitions for a specific fermion contribution by adding $f$ in the superscripts. Thanks to the exclusiveness of the LbL process~\footnote{The real photon or gluon emission contributions are zero because of the Furry's theorem and the conservation of colour and $C$ number in QCD and QED.}, we can even include partial next-to-next-to-leading order (NNLO) contributions by squaring the whole one- and two-loop amplitudes in the partonic cross-sections, that we denote as NLO$^\prime$:
\begin{equation}
\begin{aligned}
\hat{\sigma}^{{\rm NLO}^\prime_{\rm QCD+QED}}=\frac{1}{2s}\int{{\rm d}\Phi_2\overline{\sum}_{\rm helicity}{\left|\mathcal{M}_{\lambdavec}^{(0,0)}+\mathcal{M}_{\lambdavec}^{(1,1)}\right|^2}},\label{eq:xsdef2}
\end{aligned}
\end{equation}
which is the prediction we provide here.

The physical LbL cross-sections of ${\rm A}{\rm B}\overset{\gamma\gamma}{\to}{\rm A}\gamma\gamma{\rm B}$, with beam particles ${\rm A}$ and ${\rm B}$, should convolve the partonic cross-section with the corresponding photon-photon flux $\mathcal{L}^{({\rm AB})}$ so that $\sigma_{{\rm A}{\rm B}}=\int{{\rm d}x_1{\rm d}x_2\mathcal{L}^{({\rm AB})}(x_1,x_2)\hat{\sigma}}$,
where $x_1$ and $x_2$ are the longitudinal momentum fractions carried by the initial-state photons. Note that the photon-photon flux cannot simply be factorized into a product of two photon density functions of ${\rm A}$ and ${\rm B}$ in the case of UPCs~\cite{Shao:2022cly} because of the presence of a non-trivial survival probability. In the special case of ${\rm A}={\rm B}=\gamma$, we take $\mathcal{L}^{(\gamma\gamma)}(x_1,x_2)=\delta(1-x_1)\delta(1-x_2)$ with $\delta$ being the Dirac delta function.

\begin{figure}[hbt!]
\includegraphics[width=0.95\columnwidth,draft=false]{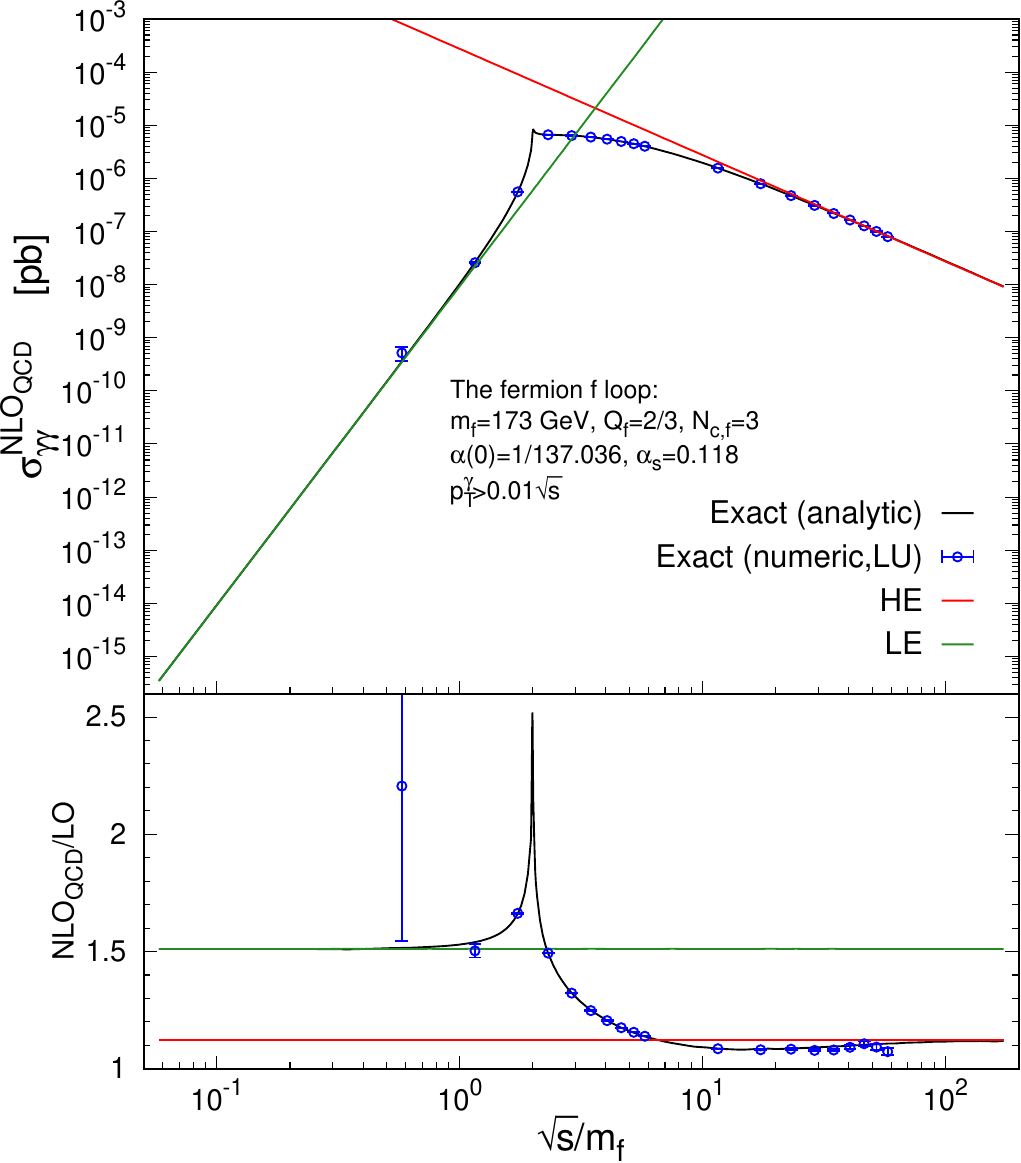}
\caption{The NLO QCD LbL cross-section (upper) as well as the $K$ factor (lower) for a given quark contribution. The black curve and the blue points denote the full mass calculations with the analytic and numeric approaches. The red and green curves represent the HE and LE approximations.}
\label{fig1} \vspace*{-0.5cm}
\end{figure}

In order to assess the mass effect, without losing generality, let us first consider the NLO QCD corrections to $\sigma_{\gamma\gamma}$ with a single massive top quark loop of $m_f=173$ GeV. We fix the $\alpha$ value in the Thomson limit $\alpha(0)=1/137.036$ and the strong coupling constant $\alpha_s=0.118$. We also impose a transverse momentum cut on the final photons $p_T^\gamma>\sqrt{s}/100$. In the center-of-mass frame of the two initial-state photons, this cut corresponds to limiting the scattering angle to be within $1.15^\circ\lesssim\theta_\gamma\lesssim 178.85^\circ$, or equivalently the pseudo-rapidity to be $|\eta^\gamma|\lesssim 4.6$. The NLO QCD cross-section in terms of $\sqrt{s}/m_f$ from the analytic amplitudes is given as the black curve in the upper panel of fig.~\ref{fig1}. We also overlay with blue points the results from the numerical LU approach, with its corresponding MC uncertainty. Each of these results with LU has been obtained using 100M sample points across all FSGs, computed in 50 CPU-hours to reach below 1\% in the NLO correction. The two approaches agree well within the errors except in the asymptotic limits $\sqrt{s}\ll m_f$ and $\sqrt{s}\gg m_f$, where numerical instabilities prohibit a fair comparison. We also compare our full mass-dependent results with the two known approximations: the high-energy (HE)~\cite{Bern:2001dg} (red) and the LE~\cite{Martin:2003gb} (green) limits. Our exact results match well with these two approximations in their applicable regimes. In particular, the relative difference between the exact computation and the HE (LE) approximation reaches below $2\%$ when $\sqrt{s}/m_f>37$ ($\sqrt{s}/m_f<0.52$). A kink appears at the threshold $\sqrt{s}\to 2m_f$, where the two-loop amplitudes suffer from the Coulomb singularity. Such a singularity is however integrable and thus harmless when convolved with the realistic photon-photon flux that exhibits a wide spectrum. A proper treatment of this region requires a Coulomb resummation, which is beyond the scope of this letter. The $K$-factors of $\sigma_{\gamma\gamma}^{{\rm NLO}_{\rm QCD}}/\sigma_{\gamma\gamma}^{{\rm LO}}$ are displayed in the lower panel of fig.~\ref{fig1}. Our exact computation exhibits a non-trivial shape of the $K$-factor as a function of the collision energy, whereas the HE and LE approximations are close to the constants $1.124$ and $1.512$ respectively. In contrast to the LE limit, the exact mass-dependent $K$-factor approaches the HE limit slowly.

We now collect all SM contributions from the charged fermions and the $W^\pm$ boson, and perform a data-theory comparison for the ATLAS Pb-Pb UPC measurement at $\sqrt{s_{\rm NN}}=5.02$ TeV~\cite{ATLAS:2020hii} with our state-of-the-art computation. Since we have renormalized the internal fermion masses in the OS scheme, we set pole masses to the following values: $m_\mu=0.1134$ GeV, $m_\tau=1.777$ GeV, $m_c=1.5$ GeV, $m_b=4.75$ GeV, $m_t=172.69$ GeV and $m_W=80.377$ GeV, whereas the remaining SM charged fermions-- electron, up, down and strange quarks --are kept massless. The EM coupling $\alpha$ associated to the external on-shell photons is set to $\alpha(0)=1/137.036$, but we consider the renormalization group running of the couplings $\alpha_s$ and $\alpha$ in the loops. For the central value of the renormalization scale, we choose $\mu_{R,0}=m_{\gamma\gamma}/2$. We then estimate missing higher order contributions by varying the renormalization scale by the factors $1/2$ and $2$. The scale evolution of the strong coupling $\alpha_s(\mu_R)$ follows the standard 2-loop modified minimal subtraction scheme with variable flavor number and $\alpha_s(m_Z)=0.118$. The running of the coupling $\alpha(\mu_R)$ is obtained from the photon propagator at momentum transfer $\mu_R$, with $\alpha(m_Z)=1/127.955$. We also considered the alternative EW $G_\mu$ renormalization scheme for $\alpha$ and verified that it yields insignificant differences.
For the Pb-Pb UPC photon-photon flux $\mathcal{L}^{({\rm PbPb})}$, we consider the charge form factor (ChFF) available in \gammaupc~\cite{Shao:2022cly}, and we tested that the resulting $K$-factors are mostly independent of the choice of flux but of course not the resulting cross sections.

\begin{figure}[hbt!]
\includegraphics[width=0.95\columnwidth,draft=false]{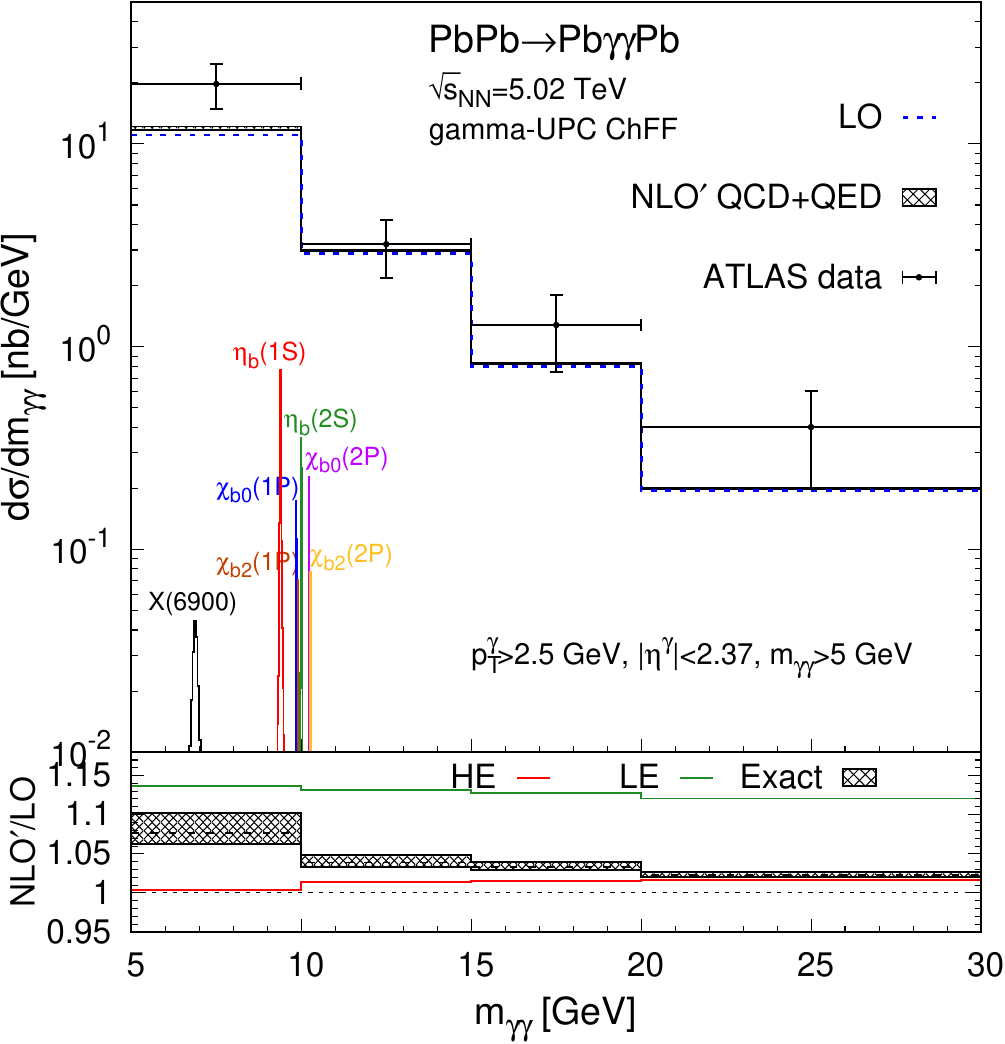}
\caption{The comparison between the LO (blue dashed), NLO$^\prime$ QCD+QED (grey hatched) with the ATLAS measurement~\cite{ATLAS:2020hii} for the di-photon invariant mass distribution. The $C$-even resonances are also displayed in the upper panel, while the $K$ factors from the two approximations are reported in the lower panel.}
\label{fig2} \vspace*{-0.5cm}
\end{figure}

The NLO$^\prime$ QCD+QED cross-section within the ATLAS fiducial volume increases by $6.5\%_{-1.2\%}^{+2.1\%}$ with respect to (wrt) the LO cross-section of $76$ nb~\footnote{The LO cross section perfectly agrees with the number quoted in ref.~\cite{Shao:2022cly} from \gammaupc+\mgamc.}.
Our final prediction of $81.2_{-0.9}^{+1.6}$ nb~\footnote{For reference, we also indicate that the pure NLO QCD+QED corrections, \ie,  excluding the partial NNLO contributions from NLO$^\prime$ prediction, enhance the LO fiducial cross-section by $5.9\%^{+2.0\%}_{-1.1\%}$ for a total cross-section $80.7^{+1.5}_{-0.8}$ nb. Moreover, the pure NLO QCD and QED corrections change the LO cross-section by $+6\%$ and $-0.1\%$ respectively.} stands $1.8\sigmaup$ below the ATLAS measurement of $120\pm 22$ nb, which is a reduction of the tension observed when comparing to LO predictions. The quoted uncertainty is obtained from a $(1/2,2)$-variation of the renormalization scale. For reference, we stress that the LE and HE approximations for the $K$-factors of the tau, charm quark and bottom quark contributions increase the LO cross-section by $13\%$ and $0.7\%$ respectively (see bottom layout of fig.~\ref{fig2}). This significantly differs from our prediction and thus highlights the importance of retaining exact fermion mass dependence in the computation of NLO corrections to the LbL cross-section.

Fig.~\ref{fig2} reveals that the tension between data and theory is largest in the first di-photon invariant mass bin $m_{\gamma\gamma}\in [5,10]$ GeV. This observation motivates the studies of the impact from the $C$-even bottomonia~\cite{Krintiras:2023axs} and the first experimentally-observed fully-charmed tetraquark state $X(6900)$~\cite{Biloshytskyi:2022dmo}. We have also simulated $6$ $C$-even bottomonium states as well as $X(6900)$ with the \ho\ event generator~\cite{Shao:2012iz,Shao:2015vga}. The inclusive cross-sections of these resonances are proportional to the square of their di-photon decay widths (see eq.~(7) in~\cite{Shao:2022cly}), which are mostly unconstrained by experiments so that we take their values from theoretical calculations~\cite{Wang:2018rjg,Biloshytskyi:2022dmo}. In particular, the $X(6900)$ Breit-Wigner peak shown in fig.~\ref{fig2} is generated using the decay width $\Gamma_{X(6900)\to \gamma\gamma}$ computed assuming the vector meson dominance hypothesis. It turns out that the contributions of these resonances to the LbL cross-section are negligible. The study in ref.~\cite{Biloshytskyi:2022dmo} reveals bridging the gap between the LO prediction for the LbL cross-section and ATLAS data using solely the $X(6900)$ resonance, one would need to increase $\Gamma_{X(6900)\to \gamma\gamma}$ by two orders of magnitude wrt their calculation assuming vector meson dominance. Our state-of-the-art NLO$^\prime$ QCD+QED result (shown as the grey hatched band in fig.~\ref{fig2}) reduces but does not eliminate this tension. It is interesting to note that the size of NLO QCD and QED corrections is largest in the lowest $m_{\gamma\gamma}$ bin and decreases from $10\%$ down to $2\%$ in the highest mass bin. This can be explained by noting that the tau, charm, and bottom mass effects are expected to be larger at lower di-photon invariant masses, as it can be seen from the HE and LE approximations of these fermion contributions shown in the lower panel of fig.~\ref{fig2}. The exact mass-dependent $K$-factor is substantially different than these approximations. Indeed, the HE approximation strongly underestimates quantum corrections for small $m_{\gamma\gamma}$, whereas the LE approximation significantly overestimates them, especially for larger values of $m_{\gamma\gamma}$. The comparison of our NLO cross-section with ATLAS data for other observables can be found in \ref{app:moredatatheory}, as well as a more detailed comparison between the results from our numerical and analytical approaches in \ref{app:LUcomp}. The additional fiducial cut on the acoplanarity distribution due to the non-zero but small virtuality of the initial photons further reduces our theoretical predictions by around $0.5\%$, which we have ignored here.

\section{\label{sec:conclu}Conclusion}

In this letter, we have carried out for the first time the computation of NLO QCD and QED contributions to the LbL cross-section, retaining exact dependence in the fermion masses. We considered two completely different and independent computational approaches that serve as a strong cross-check of our results. The first method is more traditional and results in compact analytical expressions for the two-loop helicity amplitudes, whereas the LU method is fully numerical and was introduced only recently. This stands as the first yet-unknown cross-section obtained with LU and it highlights its potential as an alternative for tackling cutting-edge multi-loop computations. We have also compared our exact result to known approximations in the high and low energy regimes, and we find that they are only satisfactory in limited regions of the phase-space that do not cover the entire range of phenomenological interest. In particular, we find that our exact result converges slowly to the high-energy limit. Therefore, we conclude that retaining exact fermion mass dependence in the computation of higher-order corrections to the LbL cross-section is mandatory for reliable predictions. Finally, we compare our final prediction to the ATLAS measurement of LbL in Pb-Pb UPCs and find that the inclusion of exact NLO QCD and QED corrections reduces, but does not eliminate, the moderate tension observed for this process.

\paragraph*{\bf Acknowledgements}
VH thanks Charalampos Anastasiou and Zeno Capatti for their insights regarding the transient singularity appearing in the LU integrand for massless internal fermions, and Ben Ruijl for his help with the symbolic treatment of the LU numerator. HSS is grateful to David d'Enterria for pointing out the impact of the acoplanarity cut on our results. This work is supported by the grants from the ERC (grant 101041109 `BOSON', grant 101043686 `LoCoMotive'), the SNSF (grant PCEFP2\_203335), the French ANR (grant ANR-20-CE31-0015 `PrecisOnium'), the French LIA FCPPL, the European Union's Horizon 2020 research and innovation program (grant 824093 STRONG-2020, EU Virtual Access `NLOAccess').

\appendix

\section{\label{app:LUcomp}Comparisons between analytical and numerical results obtained with LU}

In this section, we provide a more detailed comparison of the differential results obtained using both our analytical and numerical approaches. We turn off the convolution with the incoming photon flux and consider LbL cross-sections for three choices of fixed collision energy corresponding to distinct kinematic regimes: a) below the $2 m_f$ threshold at $m_f/\sqrt{s}=0.5767$, b) when mass effects are important at $m_f/\sqrt{s}=0.173$, and c) when the HE approximation starts becoming viable at $m_f/\sqrt{s}=0.0865$.

\begin{figure*}[hbt!]
\centering
{\includegraphics[width=0.3315\textwidth,draft=false]{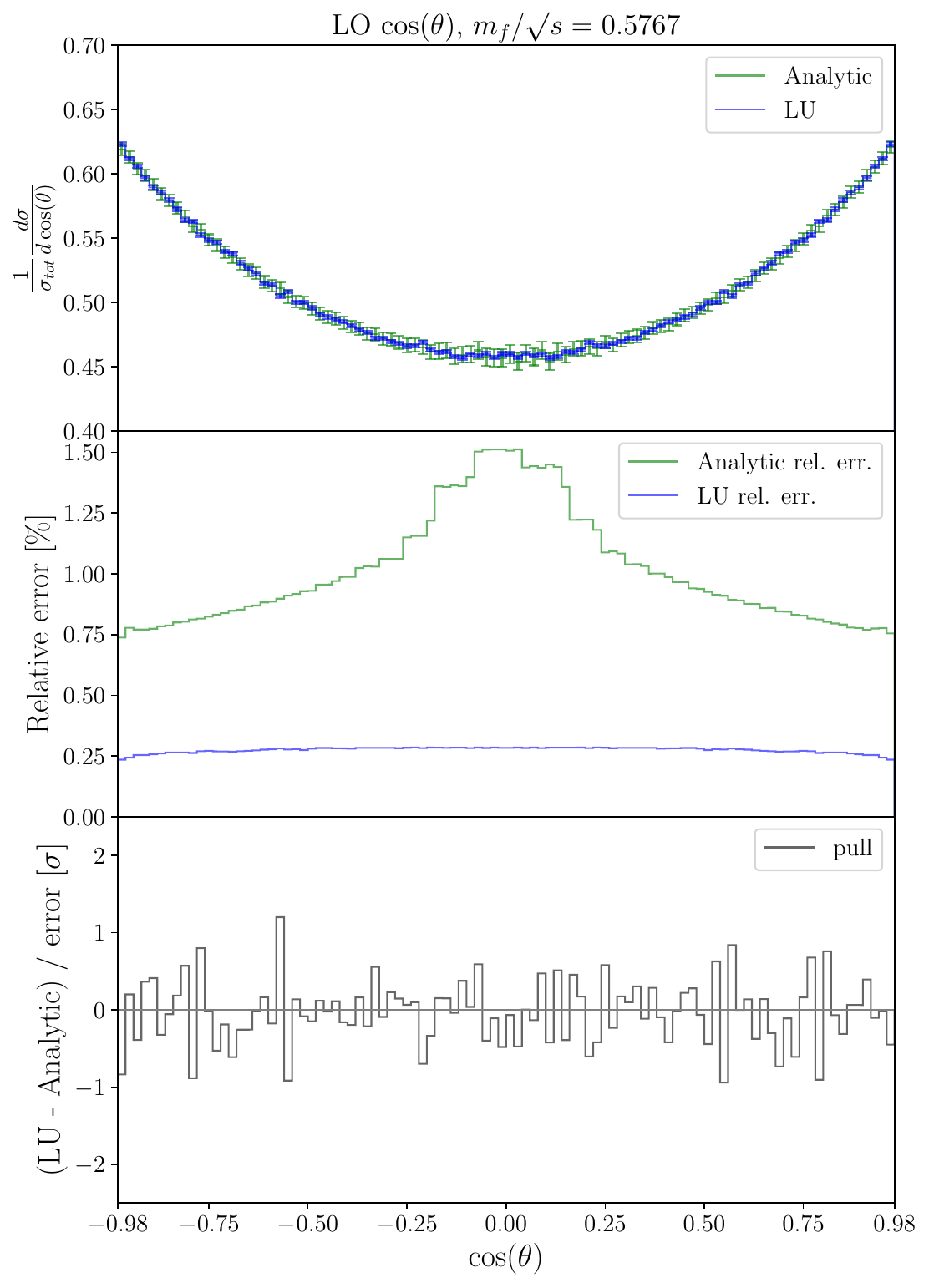}\label{diagram-a}}
{\includegraphics[width=0.327\textwidth,draft=false]{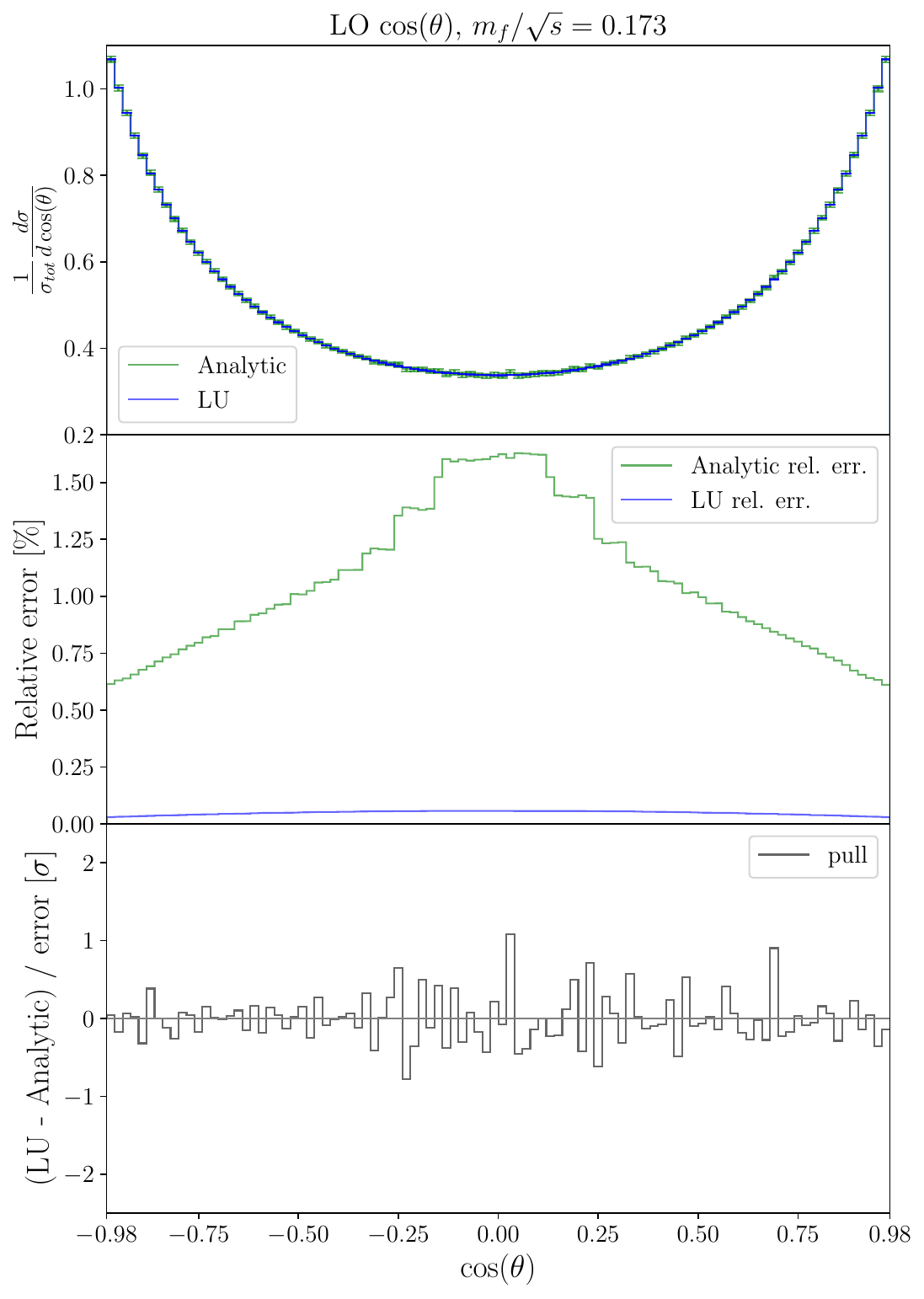}\label{diagram-b}}
{\includegraphics[width=0.3275\textwidth,draft=false]{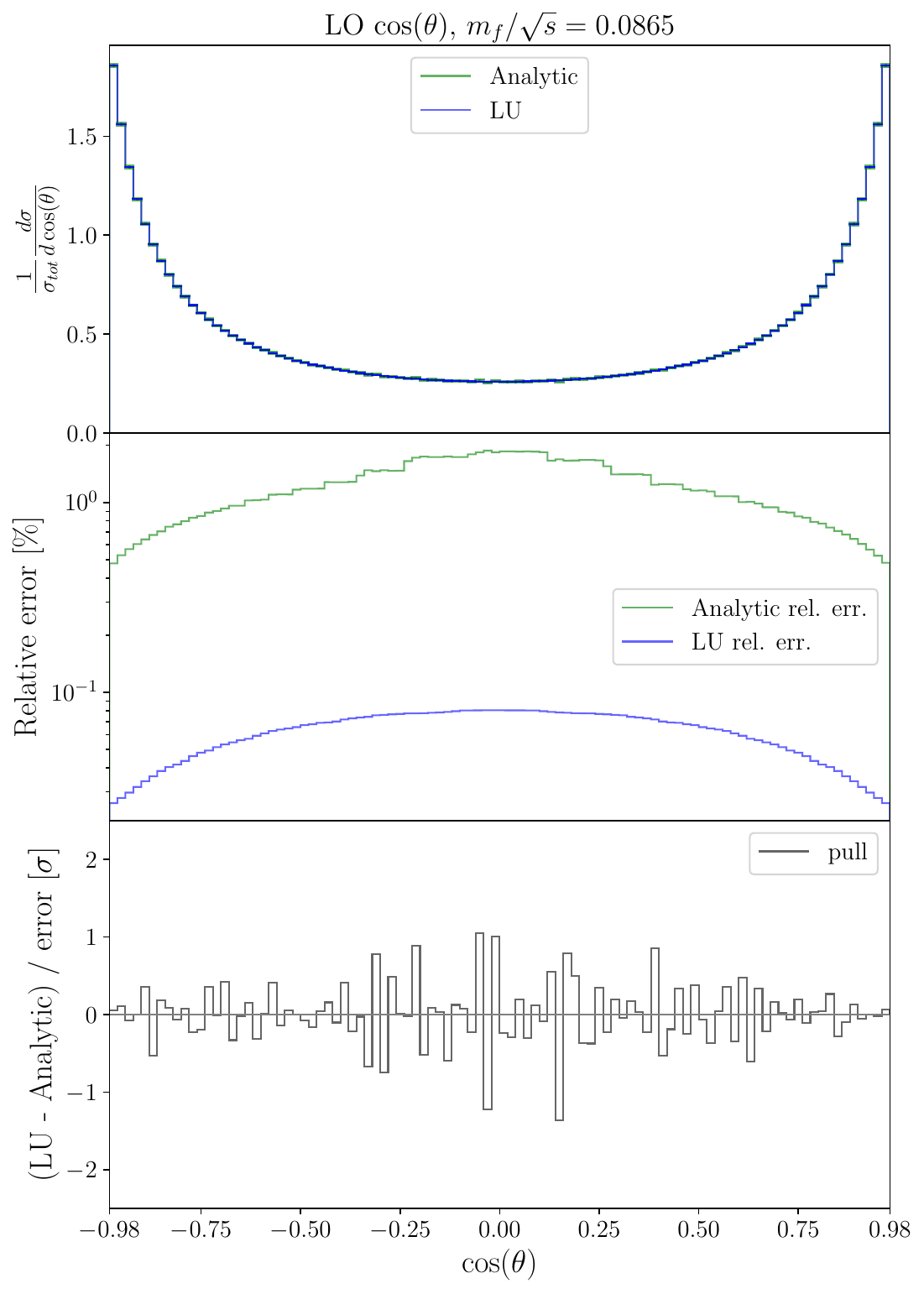}\label{diagram-c}}
{\includegraphics[width=0.325\textwidth,draft=false]{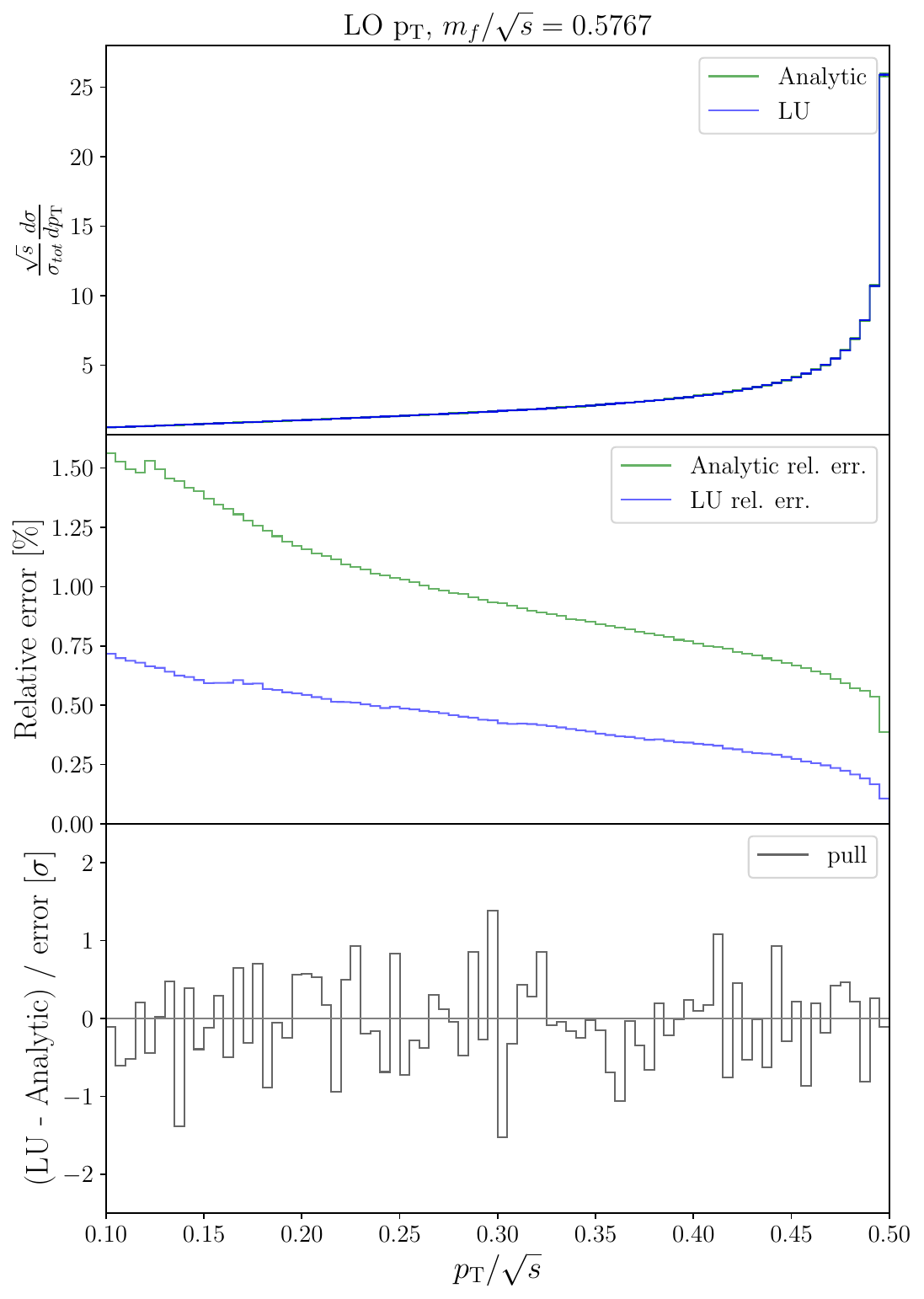}\label{diagram-a}}
{\includegraphics[width=0.3275\textwidth,draft=false]{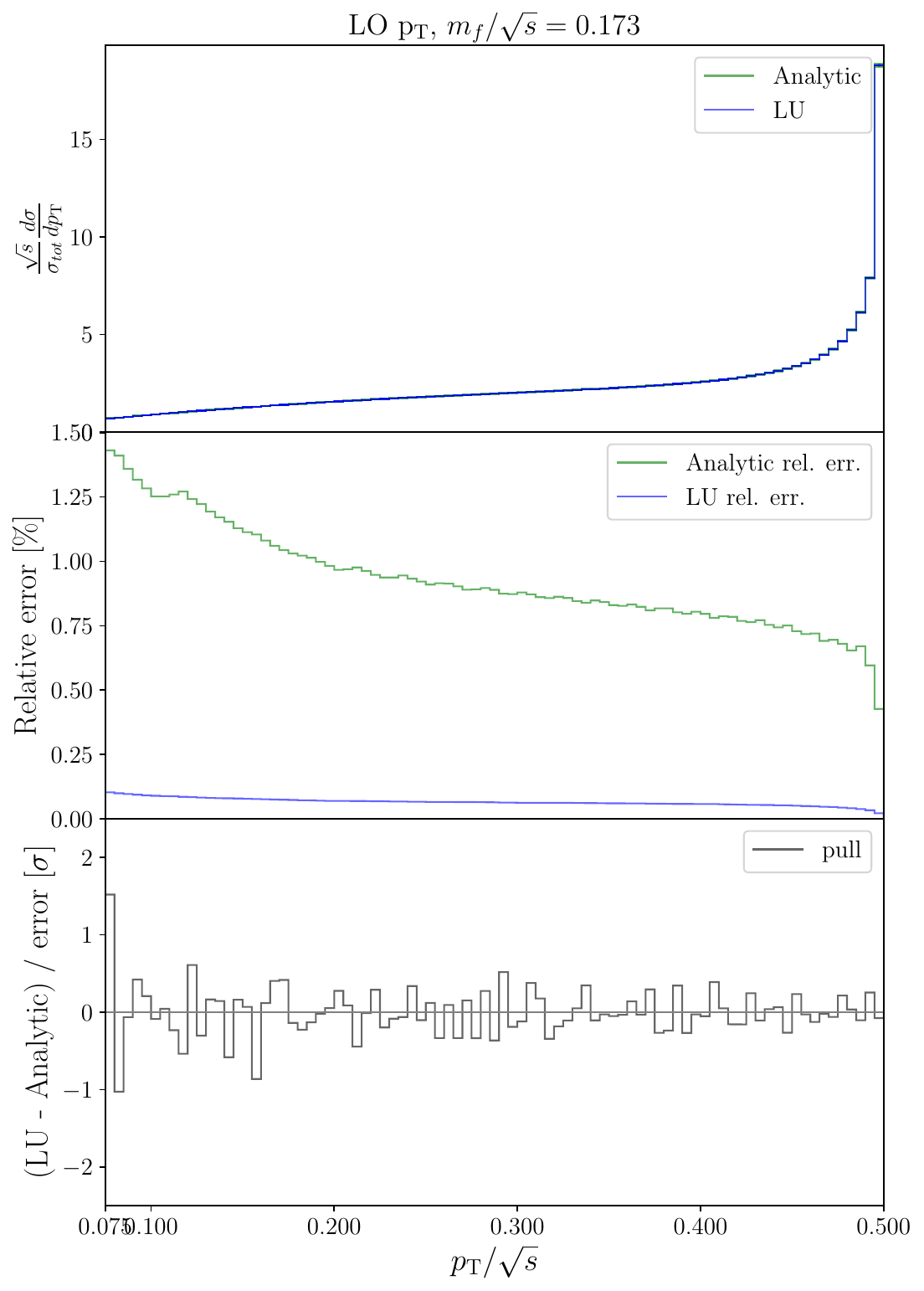}\label{diagram-b}}
{\includegraphics[width=0.33\textwidth,draft=false]{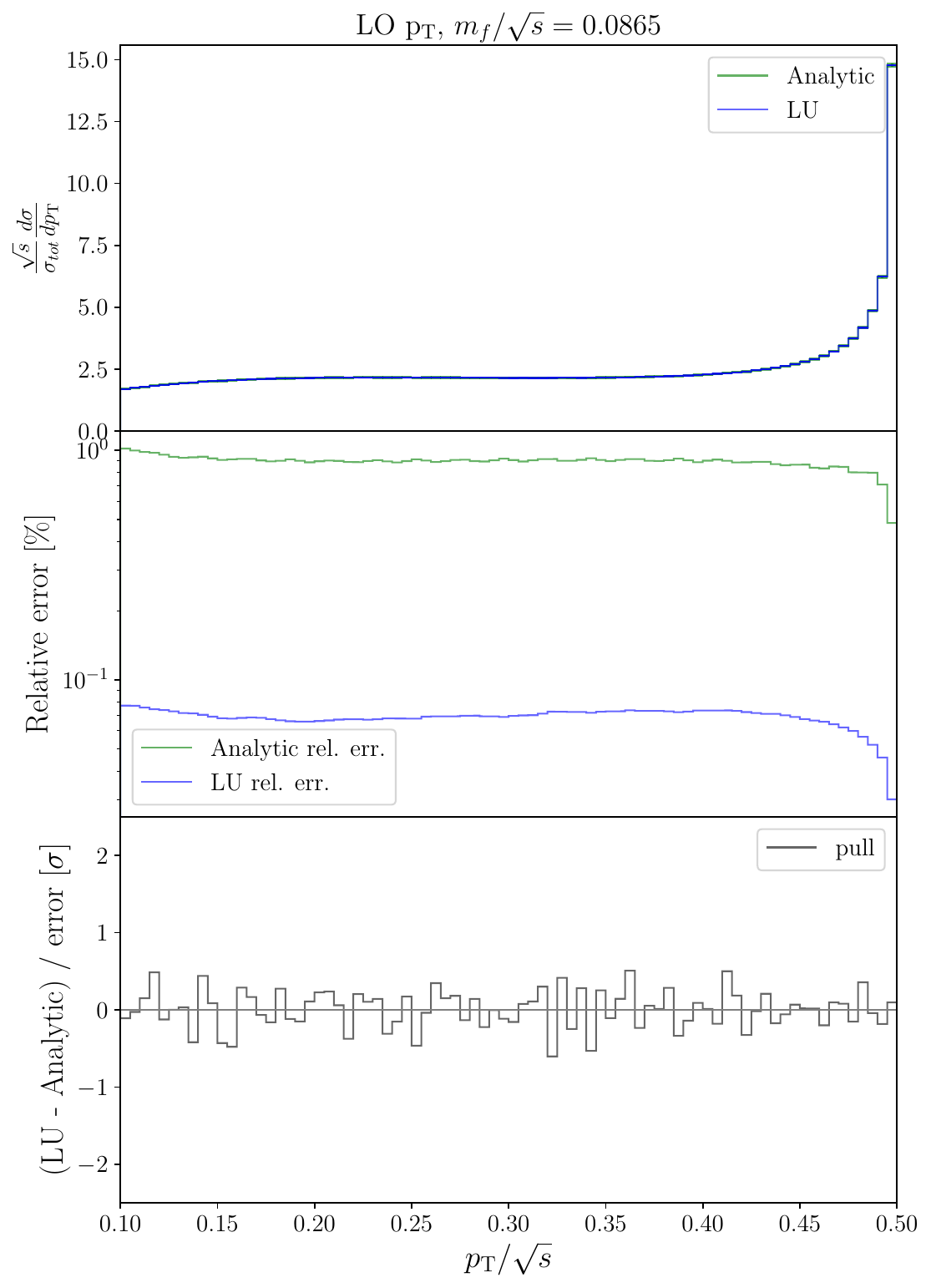}\label{diagram-c}}
\caption{Differential results for the LbL LO cross-section $\sigma_{\gamma\gamma}^{(0,0,f)}$ for the transverse momentum $p_T$ and the cosine of the angle-with-beam $\cos{\theta}$ of the final-state photons. The result using the square of the analytical one-loop amplitudes is shown in green, and the fully numerical LU result is shown in blue. The latter was obtained using 1B sample points across both contributing 2-loop FSGs, with a run time of approximately 250 CPU hours per choice of $\sqrt{s}$.}
\label{LOcomparisonFig} \vspace*{-0.5cm}
\end{figure*}

\begin{figure*}[hbt!]
\centering
{\includegraphics[width=0.33\textwidth,draft=false]{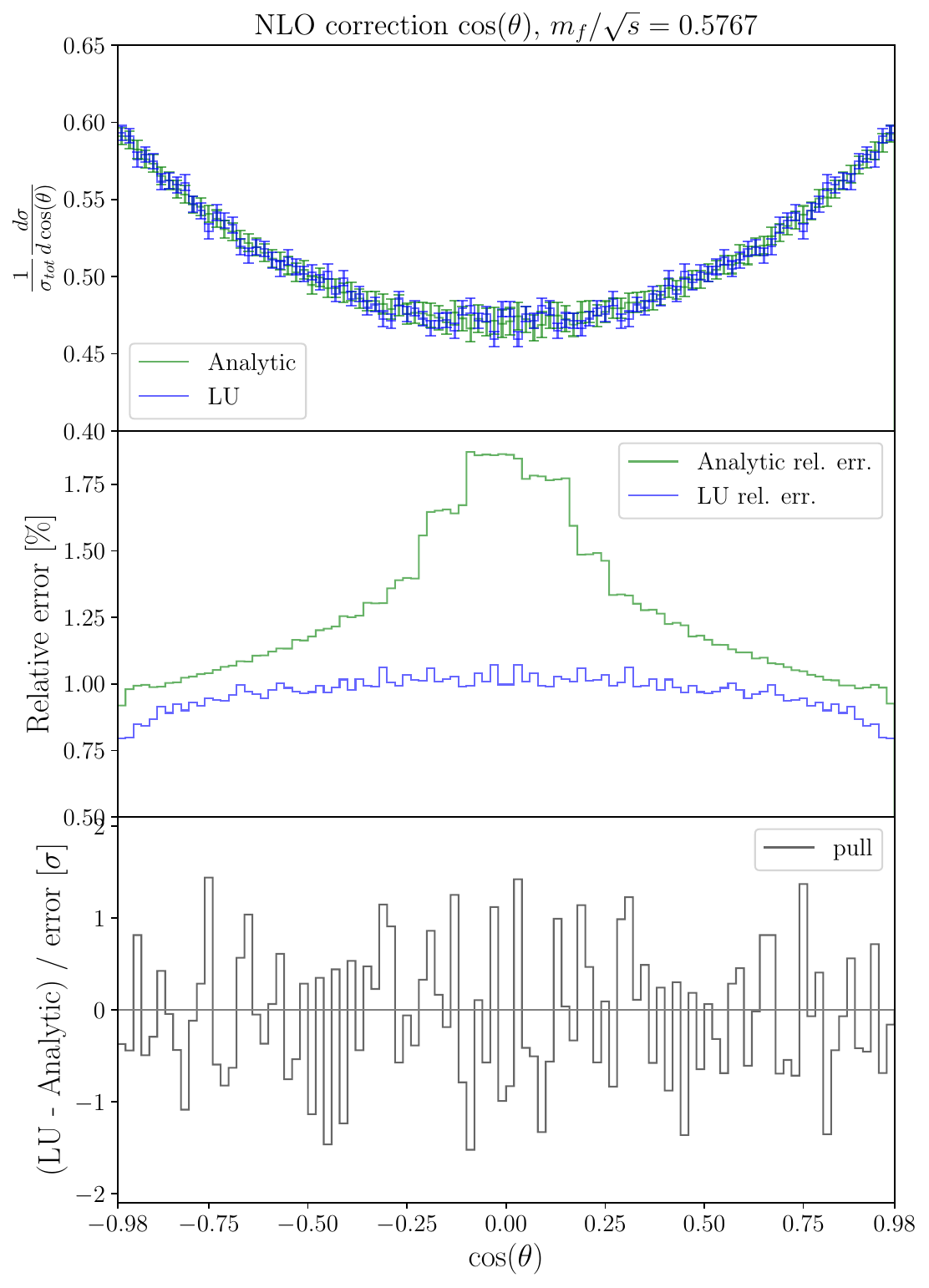}\label{diagram-a}}
{\includegraphics[width=0.33\textwidth,draft=false]{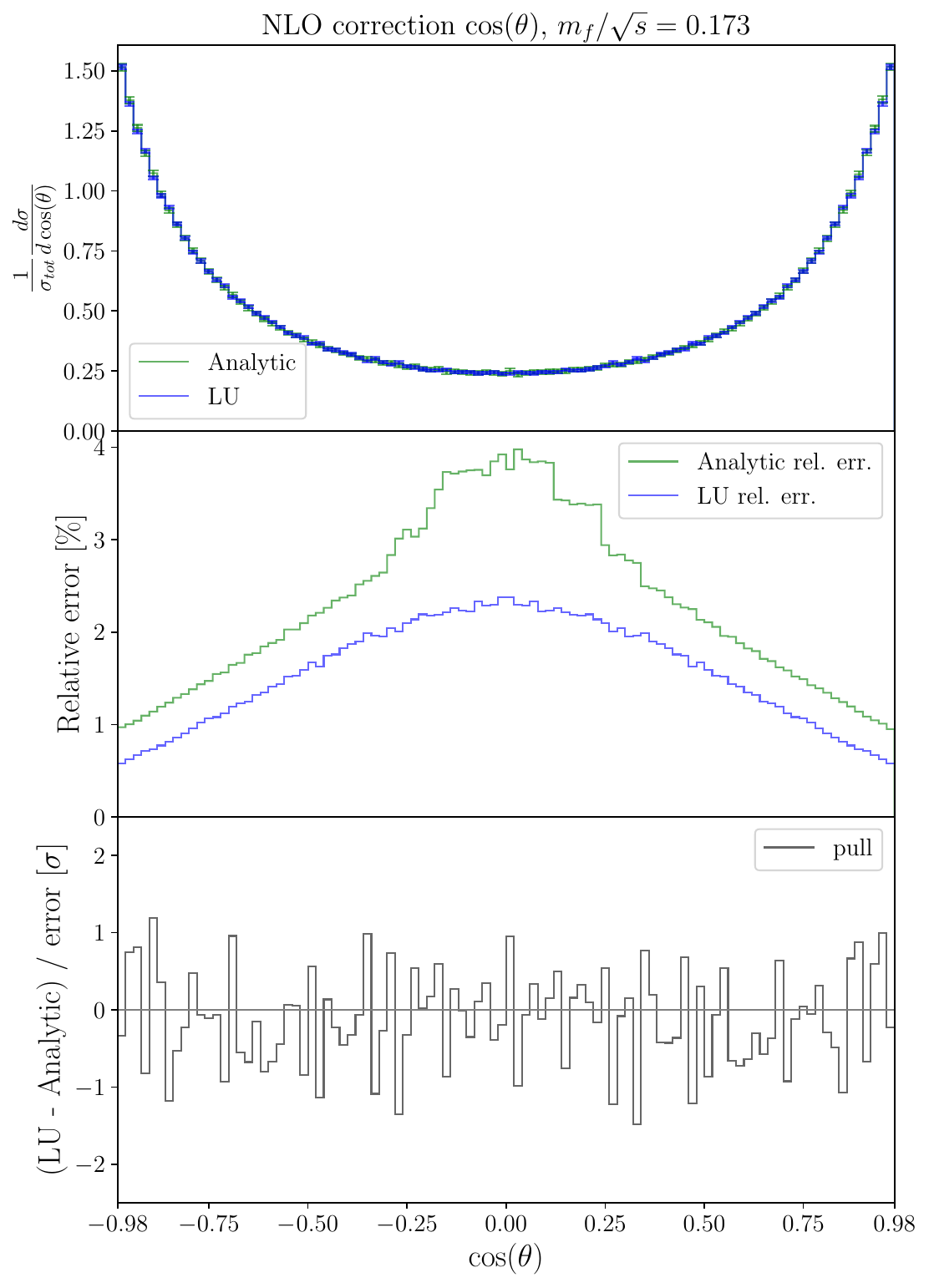}\label{diagram-b}}
{\includegraphics[width=0.3225\textwidth,draft=false]{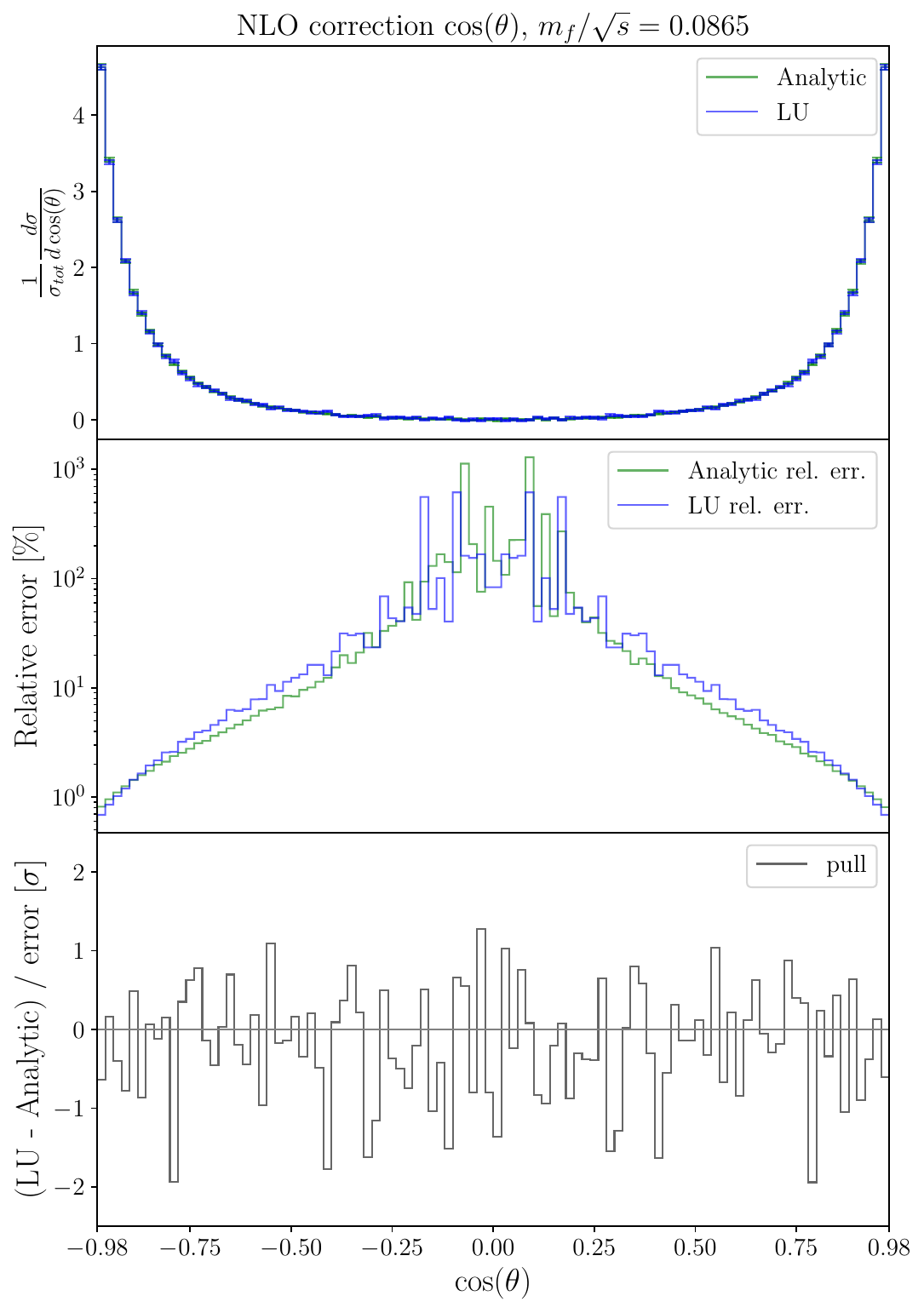}\label{diagram-c}}
{\includegraphics[width=0.32\textwidth,draft=false]{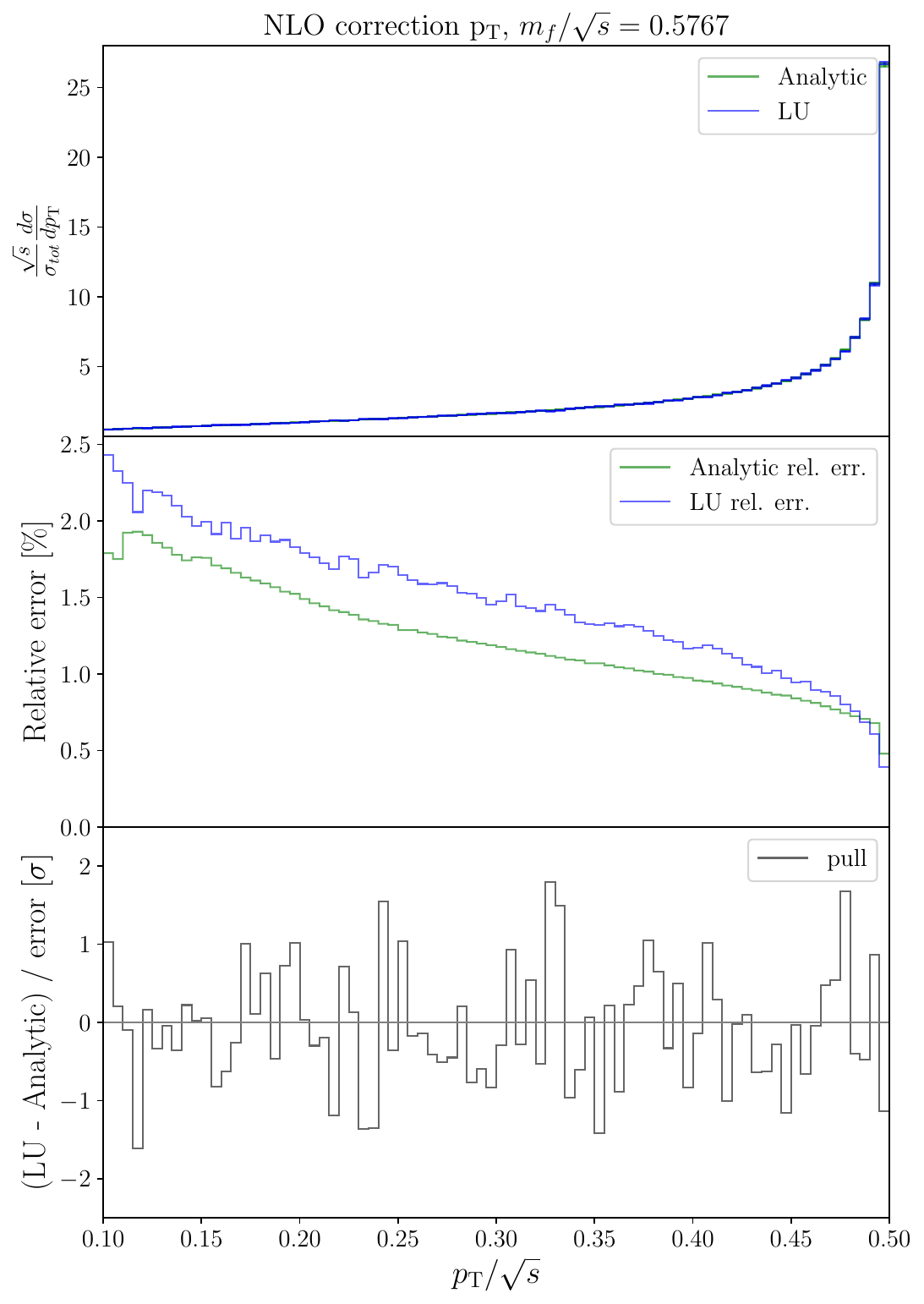}\label{diagram-a}}
{\includegraphics[width=0.33\textwidth,draft=false]{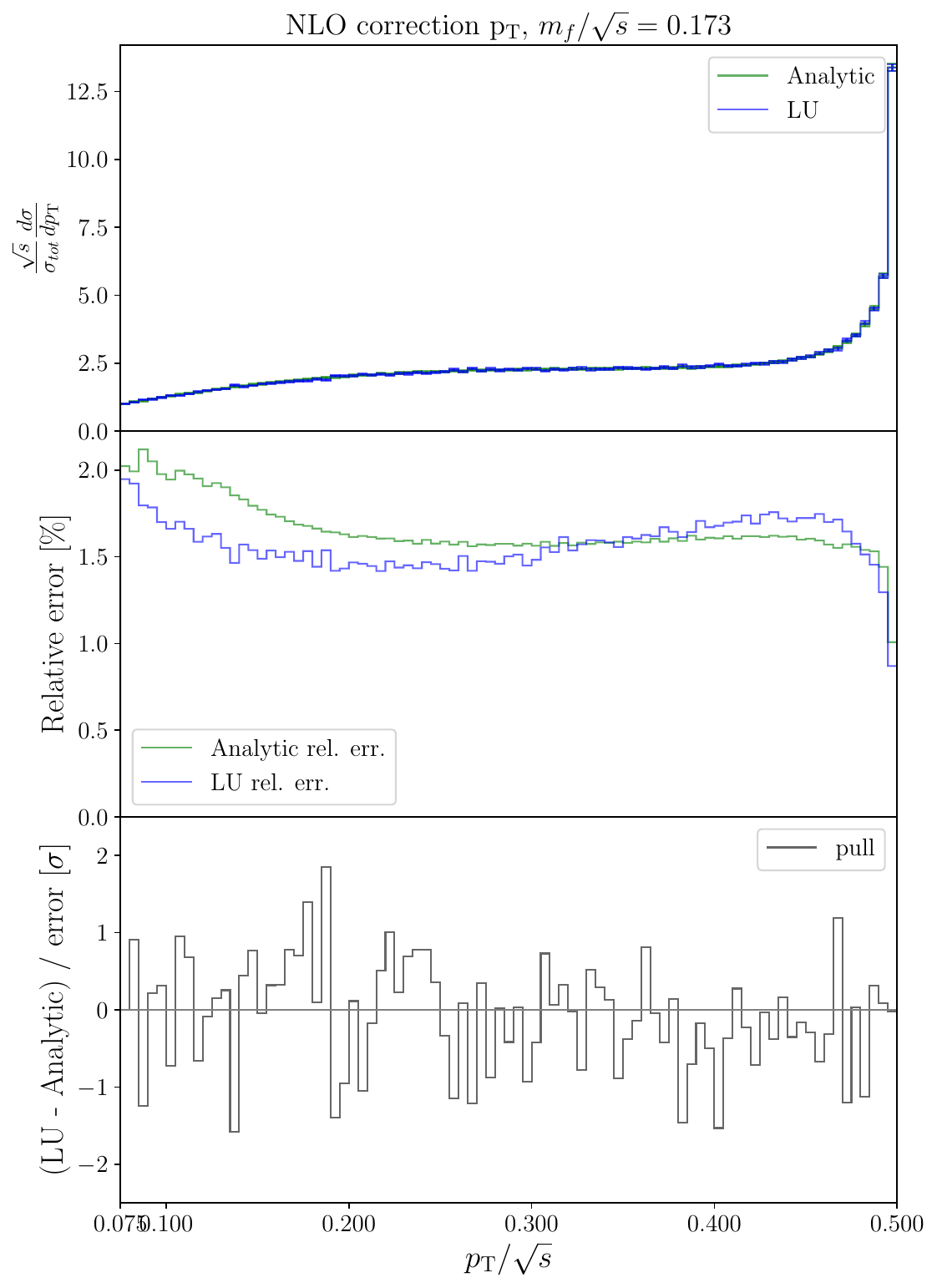}\label{diagram-b}}
{\includegraphics[width=0.3185\textwidth,draft=false]{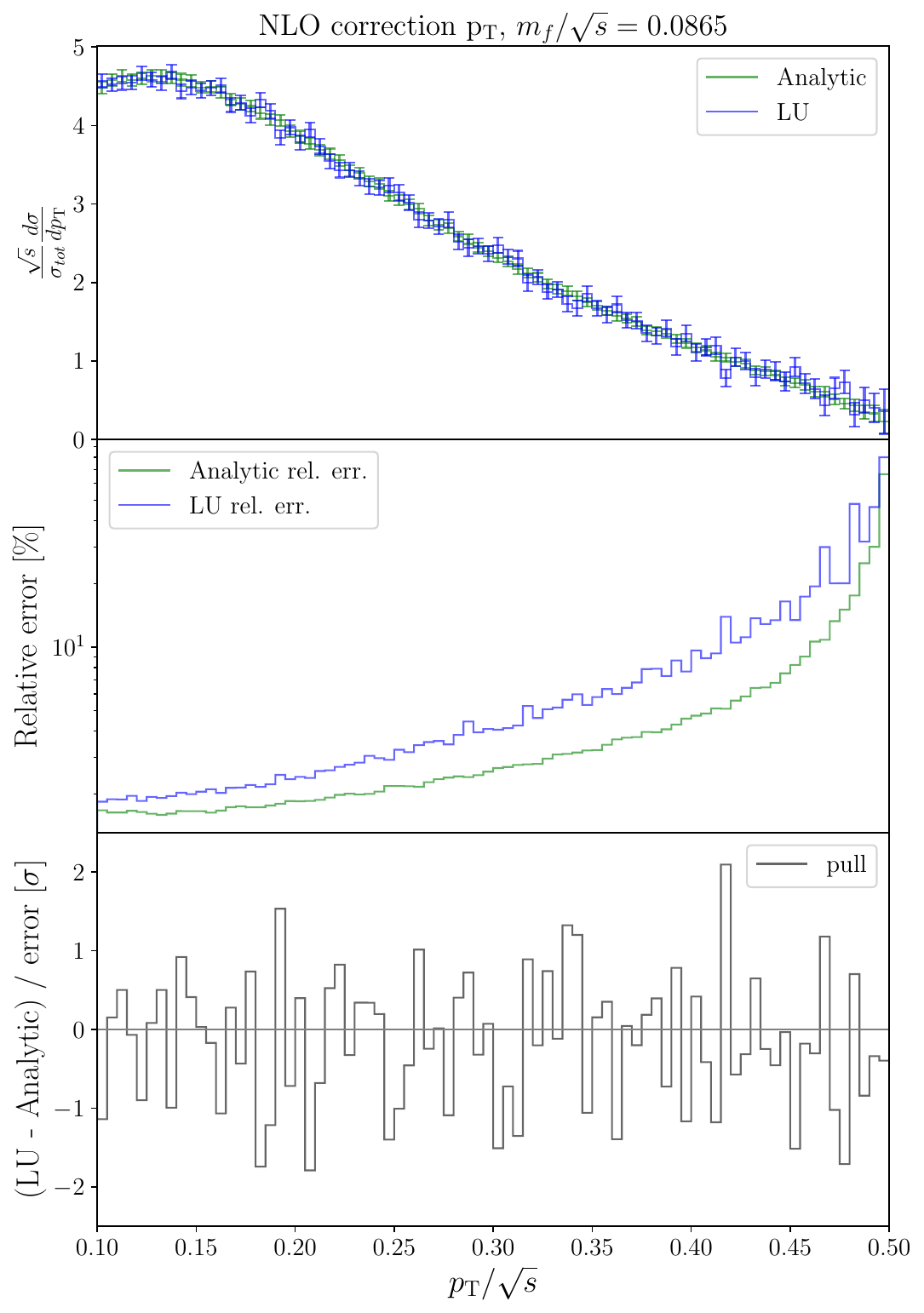}\label{diagram-c}}
\caption{Differential results for the NLO QCD correction $\sigma_{\gamma\gamma}^{(1,0,f)}$ (excl. LO contribution) to the LbL process at fixed incoming energies for the transverse momentum $p_T$ and the cosine of the angle-with-beam $\cos{\theta}$ of the final-state photons. The result using the analytical two-loop amplitudes is shown in green, and the fully numerical LU result is shown in blue. The latter was obtained using 1B sample points across all 16 contributing 3-loop FSGs, with a run time of approximately 500 CPU hours for each choice of $\sqrt{s}$.}
\label{NLOcomparisonFig} \vspace*{-0.5cm}
\end{figure*}

We consider two fully correlated observables: the transverse momentum $p_T$ of the final-state photons, and the cosine of the angle from their momenta wrt the beam $\cos{\theta}$. We report our findings in fig.~\ref{LOcomparisonFig} (LO, $\sigma_{\gamma\gamma}^{(0,0,f)}$) and fig.~\ref{NLOcomparisonFig} (NLO correction only, excl. LO contribution, $\sigma_{\gamma\gamma}^{(1,0,f)}$), and we find perfect agreement across the range displayed for both observables, with relative errors on the fiducial cross-sections at, or below, $0.1\%$ and pulls rarely exceeding $1\sigmaup$ at LO and $2\sigmaup$ for the NLO correction.

We note that numerical instabilities in our implementation of the analytic one-loop effective four-photon form factors in LU prevented us from reaching values much below $p_T/\sqrt{s} = 0.1$ (both at LO and NLO). For this reason, the $\cos{\theta}$ range was also adjusted to be within $[-0.98,0.98]$. This could easily be addressed by promoting their implementation to higher arithmetic precision.

For massless internal fermions, the numerical approach of LU suffers from a transient infrared singularity in FSG featuring internal self-energy diagrams. A transient integrand singularity is non-integrable but yields no dimensional regularization pole at the integrated level.
The origin of this transient singularity is well-understood and could be cured by either tensor-reducing the internal self-energy or including additional Cutkosky cuts traversing the massless internal fermion loop, leading to the final-state of $\gamma f\bar{f}$ which can in principle not be distinguished from $\gamma \gamma$ in the collinear limit for massless fermions.
In the absence of such a treatment, the presence of this transient singularity implies poor numerical convergence of the LU construction in the HE limit, and in practice, this renders the numerical computation impractical for $\sqrt{s}/m_f>15$. We however stress that the inclusion of the additional $\gamma f\bar{f}$ cuts is natural in light of the final experimental resolution, and that tensor-reduction of internal self-energy can be achieved with minimal effort. Analogously, the implementation of the analytic two-loop amplitudes with exact fermion mass dependence suffers from numerical inaccuracies in the LE and HE limits that limit its range of applicabilities in particular phase space corners, but which could be alleviated using higher-precision arithmetics.

Finally, we point out the stark change in shape of the photon $p_T$ distribution for $m_f/\sqrt{s}=0.0865$ wrt the LO distributions and also NLO correction distributions at lower energies.

\section{\label{app:moredatatheory}Details on the comparison of our prediction to data}

\begin{figure*}[hbt!]
\centering
\subfloat[$p_T^\gamma$]{\includegraphics[width=0.33\textwidth,draft=false]{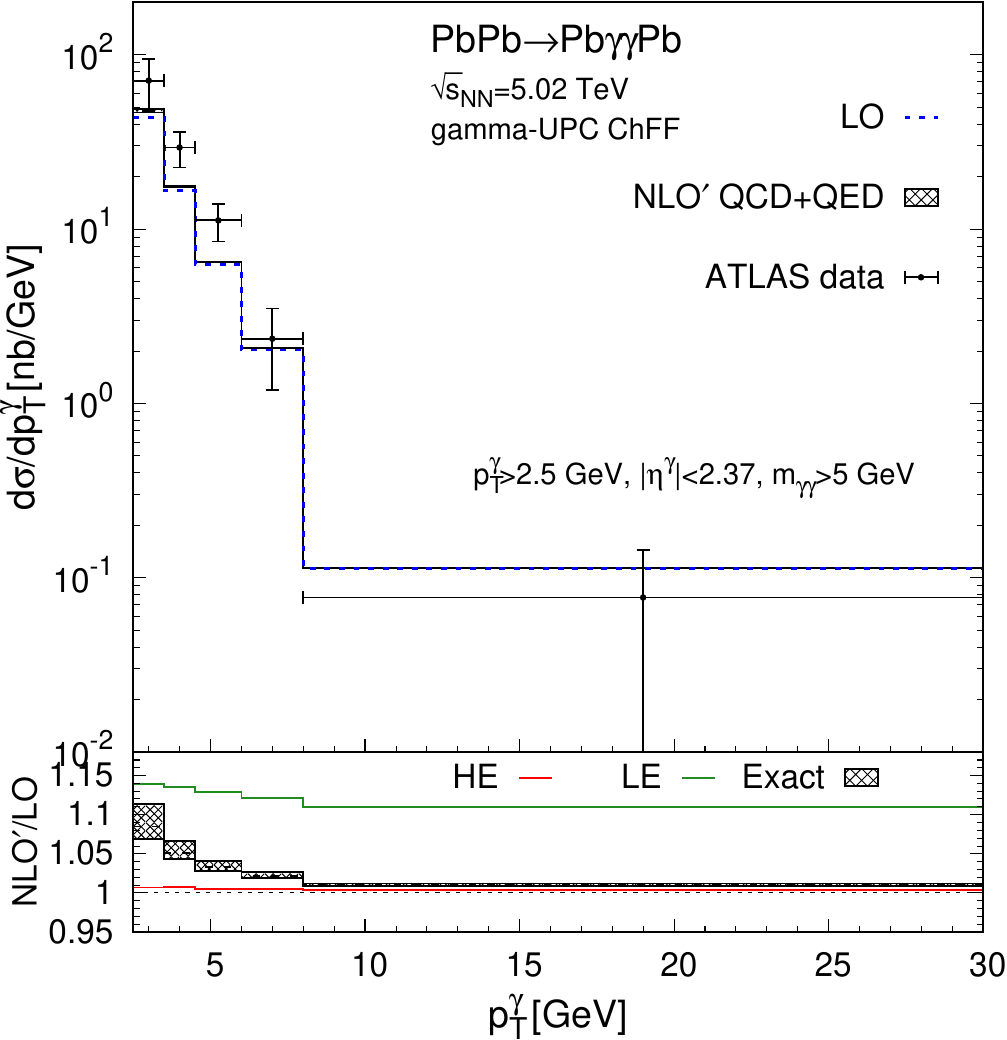}\label{diagram-a}}
\subfloat[$y_{\gamma\gamma}$]{\includegraphics[width=0.33\textwidth,draft=false]{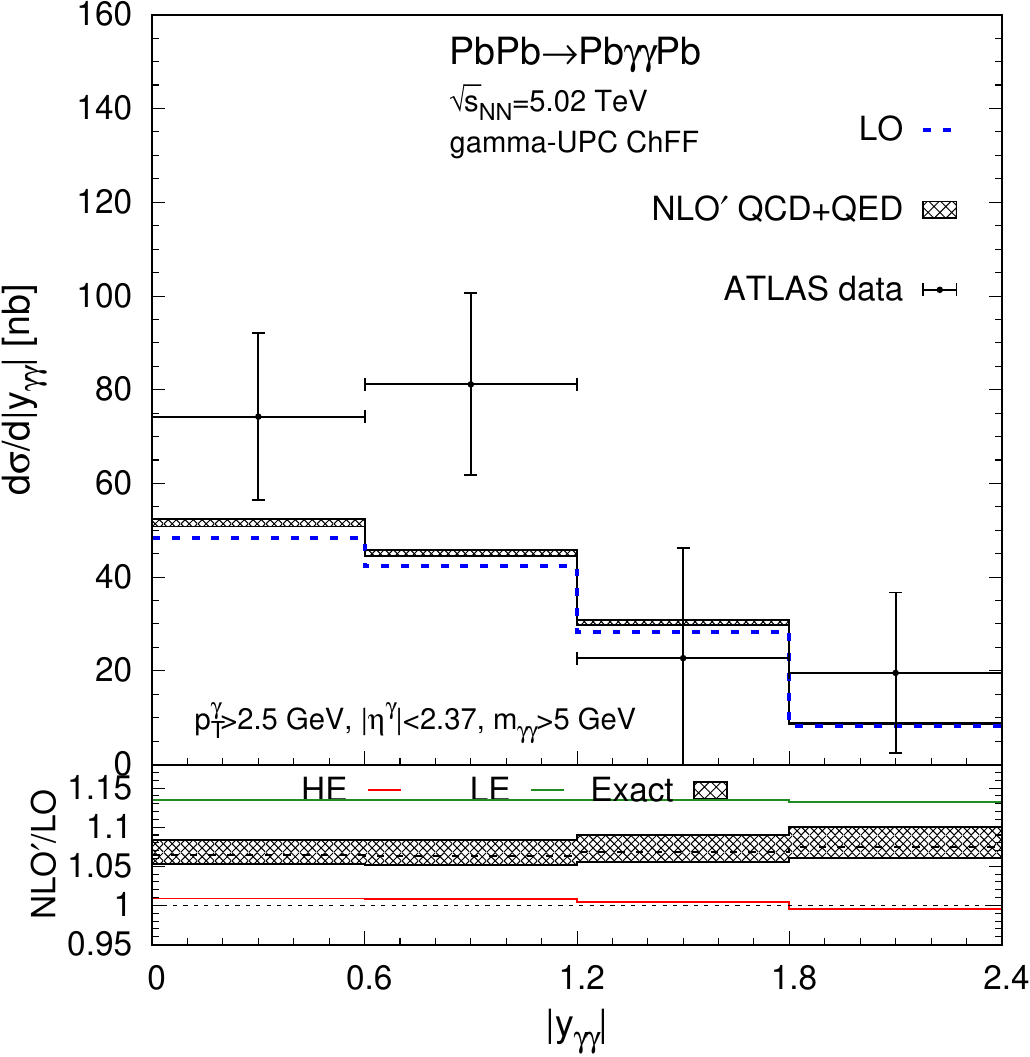}\label{diagram-b}}
\subfloat[$\cos{\theta_{\gamma\gamma}}$]{\includegraphics[width=0.33\textwidth,draft=false]{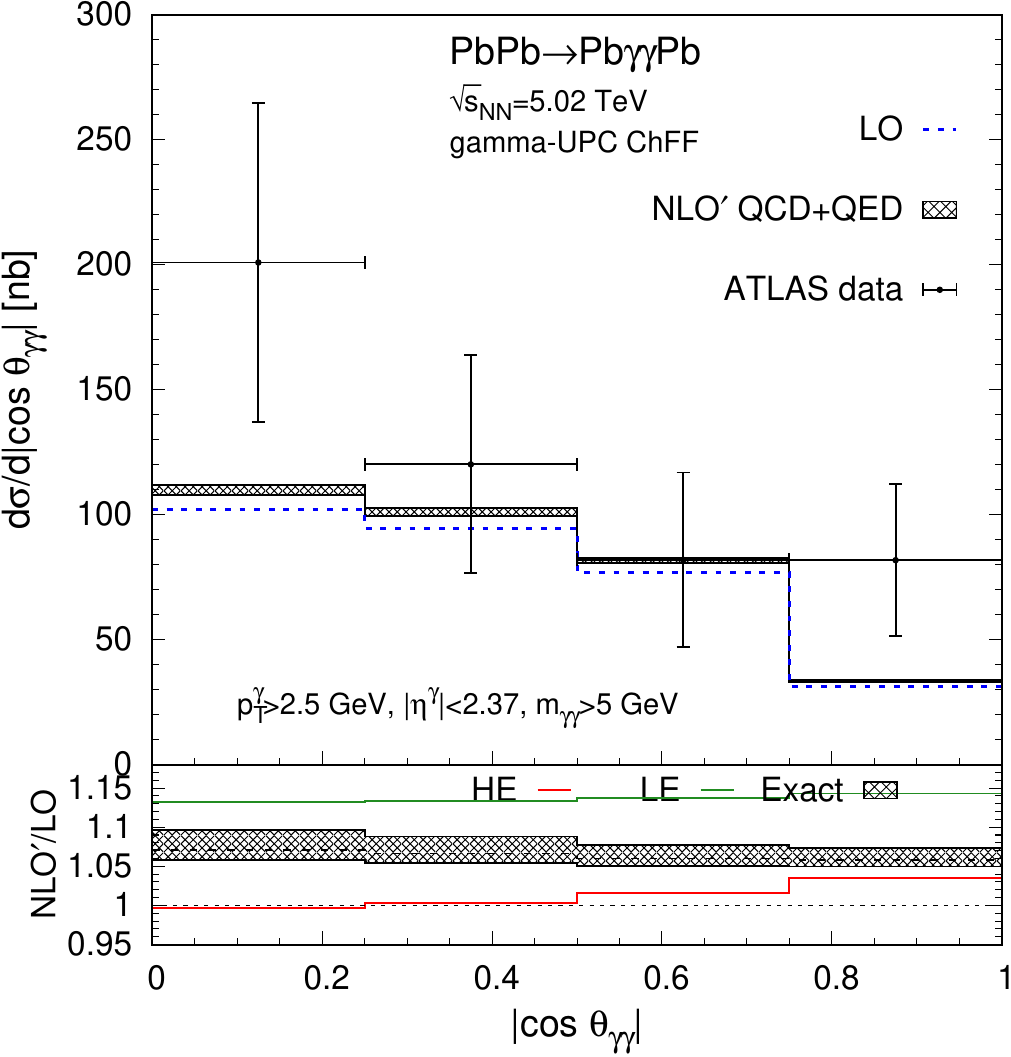}\label{diagram-c}}
\caption{The data-theory comparisons for the transverse momentum of the photon $p_T^\gamma$ (left), the rapidity of the photon pair $y_{\gamma\gamma}$ (middle) and the cosine of the scattering angle in the di-photon rest frame $\cos{\theta_{\gamma\gamma}}$ (right).}
\label{fig3} \vspace*{-0.5cm}
\end{figure*}

In this section, we present additional comparisons between our prediction and the ATLAS measurement~\cite{ATLAS:2020hii} for other observables in fig.~\ref{fig3}. From left to right, the figure shows the single photon (averaged) transverse momentum $p_T^\gamma$, the absolute value of the rapidity of the photon pair $y_{\gamma\gamma}$, and the absolute value of the cosine of the pair polar angle $\cos{\theta_{\gamma\gamma}}$. $\theta_{\gamma\gamma}$ is the scattering angle in the rest frame of the initial two photons. In each plot, the layout is similar to the one of fig.~\ref{fig2}. The $K$-factor decreases from $10\%$ at the lowest $p_T^\gamma$ value to around $1\%$ at its highest. In contrast, the $y_{\gamma\gamma}$ and $\cos{\theta_{\gamma\gamma}}$ distributions exhibit rather flat quantum corrections, however with noticeable trends. Analogous to what we observed in fig.~\ref{fig2}, the HE (LE) approximation underestimates (overestimates) the size of the exact quantum corrections in all bins.

We now briefly explain how we constructed the LE and HE approximations in figs.~\ref{fig2} and \ref{fig3} presented in this letter. In the HE limit, we take all fermions massless except for the top quark with $m_t=172.69$ GeV and the $W^\pm$ boson with $m_W=80.377$ GeV in both LO and NLO$^\prime$ QCD+QED, in order to obtain the $K$-factors. Conversely, the LE limit is obtained by setting $m_\tau=1.777$ GeV, $m_c=1.5$ GeV, $m_b=4.75$ GeV, $m_t=172.69$ GeV, $m_W=80.377$ GeV and $m_i = 0$ for $i\in\{e,u,d,s,\mu\}$. The NLO$^\prime$ LE partonic cross section is then defined to be:
\begin{equation}
\begin{aligned}
\hat{\sigma}^{{\rm NLO}^\prime_{\rm QCD+QED}}_{\rm LE}=\frac{1}{2s}\int{{\rm d}\Phi_2\overline{\sum}_{\rm helicity}{\left|\mathcal{M}_{\lambdavec}^{(0,0)}+\mathcal{M}_{\lambdavec,{\rm LE}}^{(1,1)}\right|^2}},
\end{aligned}
\end{equation}
where
\begin{equation}
\begin{aligned}
\mathcal{M}_{\lambdavec,{\rm LE}}^{(1,1)}&=&\sum_{f=e,\mu,u,d,s,t}{\mathcal{M}_{\lambdavec}^{(1,1,f)}}+\sum_{f=\tau,c,b}{\mathcal{M}_{\lambdavec}^{(0,0,f)}\frac{\mathcal{M}_{\lambdavec,{\rm LE}}^{(1,1,f)}}{\mathcal{M}_{\lambdavec,{\rm LE}}^{(0,0,f)}}}\,.  
\end{aligned}  
\end{equation}
In the above, $\mathcal{M}_{\lambdavec,{\rm LE}}^{(1,1,f)}$ and $\mathcal{M}_{\lambdavec,{\rm LE}}^{(0,0,f)}$ are the LE approximation of $\mathcal{M}_{\lambdavec}^{(1,1,f)}$ and $\mathcal{M}_{\lambdavec}^{(0,0,f)}$ respectively. It is thus clear that the LO LE partonic cross-section is identical to the exact cross-section $\hat{\sigma}^{(0,0)}$. Note that these definitions of the LE and HE limits imply that the dependence on the masses of the top quark and the $W^\pm$ boson are always accounted for exactly, although we stress that their contribution to the cross-section presented is small.

\bibliographystyle{utphys}

\bibliography{paper.bib}

\end{document}